\documentclass[10pt,journal]{IEEEtran} 

\usepackage{cite}
\usepackage{amsmath,amssymb,amsfonts}
\usepackage{algorithmic}
\usepackage{graphicx}
\usepackage{textcomp}
\usepackage{xcolor}
\usepackage{mathtools}
\def\BibTeX{{\rm B\kern-.05em{\sc i\kern-.025em b}\kern-.08em
    T\kern-.1667em\lower.7ex\hbox{E}\kern-.125emX}}

\usepackage{algorithm}
\usepackage{algorithmic}
\usepackage{comment}

\usepackage{color}
\usepackage{tablefootnote}

\usepackage{amsmath,amssymb,amsthm}
\usepackage{multirow}

\newcommand{\mbf}{\mathbf}

\newcommand{\trace}{\operatorname{trace}}

\newcommand{\nn}{\nonumber}

\usepackage{algorithm}
\usepackage{algorithmic}
\usepackage{longtable,tabularx}

\begin{document}

\title{Likelihood Scouting Via Map Inversion For A Posterior-Sampled Particle Filter}

\author{\IEEEauthorblockN{Simone Servadio,}
\IEEEauthorblockA{\textit{Iowa State University}, Ames, Iowa, USA}
\thanks{Dr. Simone Servadio is an Assistant Professor in the Department of Aerospace Engineering at Iowa State University: servadio@iastate.edu}}

% The paper headers
\markboth{Transactions on Aerospace and Electronic Systems}%
{Shell \MakeLowercase{\textit{et al.}}: Bare Demo of IEEEtran.cls for IEEE Journals}

\maketitle
\thispagestyle{plain}
\pagestyle{plain}

\begin{abstract}
An exploit of the Sequential Importance Sampling (SIS) algorithm using Differential Algebra (DA) techniques is derived to develop an efficient particle filter. The filter creates an original kind of particles, called scout particles, that bring information from the measurement noise onto the state prior probability density function. Thanks to the creation of high-order polynomial maps and their inversions, the scouting of the measurements helps the SIS algorithm identify the region of the prior more affected by the likelihood distribution. The result of the technique is two different versions of the proposed Scout Particle Filter (SPF), which identifies and delimits the region where the true posterior probability has high density in the SIS algorithm. Four different numerical applications show the benefits of the methodology both in terms of accuracy and efficiency, where the SPF is compared to other particle filters, with a particular focus on target tracking and orbit determination problems. 
\end{abstract}


\IEEEpeerreviewmaketitle

%%%%%%%%%%%%%%%%%%%%%%%%%%%%%%%%%%%%%%%%%%%%%%%%%%%%%%%%%%%%%%%%%%%%%%%%%%%%%%%%%%%%%%%%%
%%%%%%%%%%%%%%%%%%%%%%%%%%%%%%%%%%%%%%%%%%%%%%%%%%%%%%%%%%%%%%%%%%%%%%%%%%%%%%%%%%%%%%%%%
%%%%%%%%%%%%%%%%%%%%%%%%%%%%%%%%%%%%%%%%%%%%%%%%%%%%%%%%%%%%%%%%%%%%%%%%%%%%%%%%%%%%%%%%%

\section{Introduction}
\IEEEPARstart{T}{he} nonlinear filtering problem consists of estimating the state of a system affected by nonlinear dynamics using inaccurate and noisy observations. The filtering algorithm optimally estimates the state of the system by merging its own prediction of the uncertainties for the stochastic variables with the information coming from the sensors. A robust and accurate filter is fundamental to most aerospace applications, such as target tracking, orbit determination \cite{tap}, attitude estimation, spacecraft navigation, and relative pose estimation. Most, if not all, of these applications are affected by nonlinearities, which force a sub-optimal solution,  and different techniques have been developed to face the challenge. 

The Kalman filter \cite{kalman} algorithm is the correct and optimal solution only for linear and Gaussian cases, where the true posterior distribution keeps its Gaussian proprieties. Whenever nonlinearities are present, random variables undergo different approaches to propagate and approximate uncertainties. Linearization is the first proposed solution and the most common approach. The Extended Kalman Filter (EKF) \cite{gelb} linearizes the equations of motion and the measurements around the most current estimate and applies the Kalman algorithm to the linearized system. This methodology approximates probability density functions (PDF) as Gaussians, and its accuracy decays for highly nonlinear systems \cite{singla}. For such applications, a more robust uncertainty propagation technique is required. The Unscented Kalman Filter (UKF) \cite{ukf} makes use of the unscented transformation to propagate random variables under nonlinear dynamics. Thanks to the use of well-chosen sigma points, the UKF achieves higher consistency than the EKF. 

Research has been very active, especially in aerospace applications, to improve the propagation and estimation of uncertainties. In some cases, the uncertainty associated with orbital elements can be propagated analytically \cite{anal}. However, these analytical solutions are limited for detailed applications, as they do not include J2 perturbations, nor can they account for process or measurement noises. Park and Sheeres made use of State Transition Tensors (STT) to overcome linearization and propagate PDFs using their high-order central moments \cite{parksA}. Their STT propagation technique is expanded in a complete filter algorithm in \cite{parksB}, with a Kalman algorithm able to handle measurement updates and process noises. However, in \cite{parksB}, they only update the mean and covariance of the state of the system, discarding the STT information to update higher-order central moments. Afterward, Majji et al. \cite{majji} expanded the work in \cite{parksB}, introducing a tensorial mechanization to include measurement updates in all higher-order central moments. Valli et al. \cite{valli} recreated the STT mathematics using Taylor polynomials in the Differential Algebra (DA) framework, obtaining similar results to \cite{majji}. The high-order DA filter has been applied to the simple two-body dynamics in \cite{valli2} and the more challenging applications of relative pose estimation in \cite{jsr}, assessing the effectiveness and robustness of the DA technique. Other uncertainty propagation techniques based on central moments, like the use of the Koopman Operator for space applications \cite{kof}, are usually followed by classic Kalman update equations to develop complete filtering algorithms.

These mentioned techniques approximate distributions undergoing nonlinear dynamics using different methodologies, but they still create a linear estimator, i.e., the state estimate is a linear function of the measurement. Indeed, STT, DA, and UT use different approaches to predict means and covariances to get a more accurate evaluation of the Kalman gain. Still, the measurement always appears linearly in the Kalman update equation. Nonlinear estimators usually outperform linear ones. The easiest solution is the Gaussian Sum Filter (GSF), which divides the distribution into multiple Gaussian kernels (known as Gaussian Mixture Model, GMM) and applies the EKF for each model \cite{gsf}. The GSF estimate is the weighted sum of the estimates of each model according to the likelihood function. The other common approach is to achieve a nonlinear estimator by representing distributions using samples and developing particle filters such as the Bootstrap Particle Filter (BPF) \cite{bpf}. Recently, a nonlinear estimator has been achieved by designing the polynomial update \cite{recursive}, improving accuracy levels by keeping the linear update structure \cite{jaif}. The polynomial update has been developed in the DA framework by using central moments of Gaussian random variables \cite{Servadio2021a}.

The Differential Algebra Core Engine (DACE) represents variables as their Taylor polynomial expansion. This representation had significant development in space applications, especially for GN\&C problems. The polynomial representation has been helpful in propagating uncertainties \cite{cele}. The polynomial evaluation of the dynamics led to the fast propagation of an ensemble of particles \cite{jota}, providing computational advantages compared to classical numerical propagations \cite{massari}. The convergence and accuracy of the polynomials have been studied to guarantee low accuracy errors. Automatic Domain Splitting (ADS) checks for the accuracy of the Taylor expansions at the edges of the domain. It divides it into smaller sections whenever a threshold error is not met. The result is the creation of multiple polynomials to propagate large uncertainties \cite{sets} for long times. 

Sequential Monte Carlo methods, such as particle filtering, are common nonlinear estimators where PDFs are approximated as Probability Mass Function (PMF) using a set of particles \cite{seqmc}. The Bootstrap Particle Filter (BPF) \cite{bpf} propagates every particle individually and assigns weights according to the measurement likelihood. Resampling is performed whenever a small percentage of the particle acquires most of the overall weight, and heavy particles are divided into multiple samples with the same smaller weight. The Gaussian Particle Filter (GPF) \cite{gpf2} offers an alternative to resampling, as the updated PDF gets approximated as a Gaussian, and new particles are drawn from such PDF in the following step. Sequential Importance Samplings (SIS) help draw particles in the proximity of high-density probability regions in order to achieve an accurate estimate. The SIS is based on the selection of an importance distribution to sample from. The Unscented Particle Filter (UPF) \cite{upf} uses the UKF solution inside the SIS algorithm, adding an additional step to the previously mentioned UKF. Similarly, the Monte Carlo Filter Particle Filter (MCFPF) \cite{mcfpf} exploits the SIS mathematics using the Monte Carlo prediction of the uncertainties as importance distribution. Lastly, the Auxiliary Particle Filter (APF) uses an intermediate variable to account for the value of the measurement in the SIS importance distribution. 

The selection of the best proposal distribution has been the subject of multiple studies, especially to prove the robustness of the technique \cite{rob}. Morzfeld et al. iteratively compute a new importance distribution at the same update over the mean and covariance matrix of Gaussians\cite{imp_iter}. The use of DA has been merged with SIS already in \cite{losacco}, focusing on propagation rather than measurement updates, contrary to the developments proposed in this paper. Losacco et al. \cite{losacco} utilize ADS to define the importance distribution. Using DA polynomial evaluation, \cite{losacco} propagates states and checks accuracy and probability conditions. ADS iteratively splits the domain in two whenever the DA propagation is less accurate. The result is a prior domain split into multiple subregions of different widths, each one with its connected polynomial expansion. Importance sampling uniformly draws from each subdomain, meaning more particles are drawn for smaller subdomains.

This paper proposes a novel selection for the importance distribution that is based on transferring knowledge from the measurement back onto the state PDF. A polynomial map that connects the state space domain to the measurement codomain can be created using DA techniques. The main idea is to invert such a map to analyze the influence of the measurement noise onto the prior distribution. This can be done by the creation of scout particles that move information, creating the Scout Particle Filter (SPF). The use of DA in nonlinear filtering can be appreciated in \cite{thesis}. Other than the main map inversion in the DA framework, the paper presents a variety of novelties, such as (i) the introduction of SIS with the noise DA approximation, (ii) the double propagation in the measurement space of the measurement map, (iii) a new likelihood examination, (iv) the two separate kinds of importance distribution evaluation, (v) the underlining of a region of high posterior probability, (vi) the update decision making based on covariance examination between SPF and BPF, (vii) the definition of a new type of particle (scout), (viii) the consequent analysis of the reduced required number of particles for an accurate update, (ix) the introduction of fictitious predicted covariance approximation for a generic function, and (x) the use of the technique to bimodal posteriors where the map is inverted with bifurcation.

The paper is structured as follows. Section \ref{DA} gives an introduction to Differential Algebra and explains the creation of polynomial maps. Section \ref{SISsec} describes the Sequential Importance Samplings mathematics, and how it can be exploited to improve estimation. Afterward, the Scout Particle Filter is derived methodically in Section \ref{SPFsec}, providing assumptions, algorithm breakdown, and the system architecture. Section \ref{appsec} tests the newly developed filter against other particle filters in five different numerical applications, each time focusing on a different feature. Lastly, conclusions are drawn in Section \ref{concl}.

%%%%%%%%%%%%%%%%%%%%%%%%%%%%%%%%%%%%%%%%%%%%%%%%%%%%%%%%%%%%%%%%%%%%%%%%%%%%%%%%%%%%%%%%%
%%%%%%%%%%%%%%%%%%%%%%%%%%%%%%%%%%%%%%%%%%%%%%%%%%%%%%%%%%%%%%%%%%%%%%%%%%%%%%%%%%%%%%%%%
%%%%%%%%%%%%%%%%%%%%%%%%%%%%%%%%%%%%%%%%%%%%%%%%%%%%%%%%%%%%%%%%%%%%%%%%%%%%%%%%%%%%%%%%%
\section{The Use of Differential Algebra For Deviation Mapping} \label{DA}
Differential algebra (DA) techniques have gained increasing attention in estimation in the last years \cite{jota,Massari1}. As different filters make use of various approximations of the nonlinear transformation of random vectors, DA represents nonlinear functions with their high-order Taylor series expansion up to a selected arbitrary order. Therefore, transformations of stochastic variables are approximated by the polynomial transformation of random vectors. This representation is carried out in a new computer environment: while functions are usually based on simple evaluation at specific points in the classical floating point (FP) representation, they are represented as a matrix of coefficients and exponents in the DA framework. Therefore, any differentiable function is approximated around a specific center, and any algebraic operation, including differentiation and integration operators, is conducted as a DA structure. DA has been successfully 
used to evaluate the time evolution of the state of a dynamical system both in discrete-time applications \cite{Servadio2021a} and in continuous-time scenarios \cite{Massari1}.

For the methodology offered in this paper, we will make use of the polynomial map inversion capability of the Differential Algebra Core Engine (DACE2.0) \cite{rasotto}. Given a nonlinear function $\mbf g(\cdot)$ that transforms $\mbf y = \mbf g(\mbf x)$, the DA representation creates the Taylor polynomial mapping of deviations from the selected center $\bar{ \mbf x}$ from the domain of variable $\mbf x$ to the codomain of variable $\mbf y$. For the second-order case 
\begin{align}
    \mbf y &\approx \mbf g (\bar{\mbf x})  + D\mbf g(\bar{\mbf x}) (\mbf x -  \bar{\mbf x}) + \dfrac{1}{2} (\mbf x -  \bar{\mbf x})^TH\mbf g(\bar{\mbf x}) (\mbf x -  \bar{\mbf x}) \nn \\
    &= \bar{\mbf y} + D\mbf g(\bar{\mbf x}) \delta \mbf x + \dfrac{1}{2} \delta \mbf x^TH\mbf g(\bar{\mbf x}) \delta \mbf x\\
    &= \bar{\mbf y} + \mathcal M_{\bar{\mbf x}}( \delta \mbf x) \label{mapex}
\end{align}
where $D\mbf g(\bar{\mbf x})$ is the matrix of partial derivatives, $H\mbf g(\bar{\mbf x})$ is the Hessian matrix of $\mbf g(\cdot)$ (with $H\mbf g(\bar{\mbf x}) = DD\mbf g(\bar{\mbf x})$). The polynomial map is represented as $\mathcal M_{\bar{\mbf x}}( \delta \mbf x)$ and it maps deviations $\delta \mbf x = \mbf x -  \bar{\mbf x}$ to the space of variable $\mbf y$. Indeed, after defining $\bar{\mbf y} = \mbf g (\bar{\mbf x})$ and moving it to left hand side, Eq. \eqref{mapex} can be rewritten as
\begin{equation}
    \delta \mbf y = \mathcal M_{\bar{\mbf x}}( \delta \mbf x)
\end{equation}
These polynomials can be inverted to obtain the connection between deviations in $\mbf x$ as a function of deviations in $\mbf y$:
\begin{equation} 
    \delta \mbf x = \big( \mathcal M_{\bar{\mbf x}}( \delta \mbf x)\big)^{-1} = \mathcal W_{\bar{\mbf y}}( \delta \mbf y)
\end{equation}
which is expressed as a new polynomial map $\mathcal W_{\bar{\mbf y}}( \delta \mbf y)$ centered in $\bar{\mbf y}$.

For a more complete and detailed explanation of Differential Algebra, its recent use in space navigation and estimation, and its capabilities, please refer to the references \cite{rasotto,jass,jsr}. In particular, the polynomial evaluation replacement of numerical integration has been proven to reduce the computational burden of the software drastically, as initially shown by Berz in \cite{berz}, implemented in the original Differential Algebra library COSY INFINITY \cite{cosy}, and then perfected in the DACE \cite{massari,cele}.

%%%%%%%%%%%%%%%%%%%%%%%%%%%%%%%%%%%%%%%%%%%%%%%%%%%%%%%%%%%%%%%%%%%%%%%%%%%%%%%%%%%%%%%%%
%%%%%%%%%%%%%%%%%%%%%%%%%%%%%%%%%%%%%%%%%%%%%%%%%%%%%%%%%%%%%%%%%%%%%%%%%%%%%%%%%%%%%%%%%
%%%%%%%%%%%%%%%%%%%%%%%%%%%%%%%%%%%%%%%%%%%%%%%%%%%%%%%%%%%%%%%%%%%%%%%%%%%%%%%%%%%%%%%%%
\section{Importance Sampling} \label{SISsec}
The main idea behind particle filtering is to approximate a continuous random variable as discrete. That is, the probability density function (PDF) $p(\mbf x)$ is represented as a probability mass function (PMF), where the integral is approximated arbitrarily well with a finite sum of realizations:
\begin{align}
    p(\mbf x) &= \int_{\mathcal S(\mbf x)} p(\boldsymbol \xi)\delta(\mbf x - \boldsymbol \xi) d\boldsymbol \xi \\
    &\approx \sum_{i = 1}^N w^{(i)} \delta(\mbf x,\mbf  x^{(i)})
\end{align}
where $\delta$ is the Dirac delta function, $\mathcal S(\mbf x)$ is the support in which the PDF is defined, and $w^{(i)}$ is a non-negative weight associated with the $\mbf x^{(i)}$ realization, for a total of $N$ samples. In Sequential Monte Carlo Methods, the dynamics are usually offered in their discretized version. Thus, consider the following equations of motion, describing a nonlinear dynamical system, $\mbf f (\cdot)$, affected by a zero-mean process noise $\boldsymbol \nu$: 
\begin{subequations}\label{system}
\begin{align}
    \mbf x_{k+1} &= \mbf f_k (\mbf x_k, \boldsymbol \nu_k) \\
    \mbf y_{k+1} &= \mbf h_{k+1} (\mbf x_{k+1}) + \boldsymbol \mu_{k+1}
\end{align}
\end{subequations}
where the state of the system $\mbf x$ is assumed to have $n$ dimensions. Information about the state of the system at a given time-step $k$ is inferred by a set of measurements $\mbf y$, whose (nonlinear) function with the state, $\mbf h (\cdot)$, is assumed to be known. However, these observations are affected by a zero-mean measurement noise $\boldsymbol \mu$, such that the goal of the estimator is to provide the most accurate state by \textit{filtering out} the noise; thus the name ``filter".

In a Bayesian sense, the solution is given by propagating the PDF using the Chapman-Kolmogorov equation (CKE) \cite{barsha}. Let express with capital letters all random vectors up to the current time that is
\begin{align}
    \mbf X_k = [\mbf x_0, \mbf x_1, \dots, \mbf x_k]\\
    \mbf Y_k = [\mbf y_0, \mbf y_1, \dots, \mbf y_k].
\end{align}
Assuming that the conditional distribution is a Markovian process, i.e. $p(\mbf x_{k+1}| \mbf X_k) = p(\mbf x_{k+1}| \mbf x_k)$, and that the measurement noise is completely independent from the process noise, i.e. $p(\mbf x_{k+1}| \mbf x_k, \mbf Y_k) = p(\mbf x_{k+1}| \mbf x_k )$, the propagated distribution is therefore expressed as 
\begin{equation}
    p(\mbf x_{k+1}, \mbf x_k | \mbf Y_k) = p(\mbf x_{k+1}| \mbf x_k) p(\mbf x_k | \mbf Y_k)
\end{equation}
and the CKE is obtained from marginalization:
\begin{equation}
    p(\mbf x_{k+1} | \mbf Y_k) = \int_{\mathcal S(\mbf x)} p(\mbf x_{k+1}| \mbf x_k) p(\mbf x_k | \mbf Y_k) d\mbf x_k.
\end{equation}
Using Bayes' formulation, it is possible to calculate the posterior distribution by applying the measurement update: 
\begin{equation}
     p(\mbf x_{k} | \mbf Y_k) = \dfrac{p(\mbf y_k | \mbf x_k, \mbf Y_{k-1})p(\mbf x_k| \mbf Y_{k-1}) }{\int_{\mathcal S(\mbf x)} p(\mbf y_k | \mbf x_k, \mbf Y_{k-1})p(\mbf x_k| \mbf Y_{k-1}) d\mbf x_k}
\end{equation}
and, since the measurement noise is assumed to be independent at each time step from the other error sources, the equality for the likelihood $p(\mbf y_k | \mbf x_k, \mbf Y_{k-1})=p(\mbf y_k | \mbf x_k)$ holds. 

Working with PMF, integrals get replaced by sums since the PDF is approximated as discrete. The recursive equations that represent the evolution of the discrete PDF start after defining the initial distribution:
\begin{equation}
    p(\mbf x_0) = p(\mbf X_0,\mbf Y_0) = \sum_{i=1}^N w_0^{(i)}\delta(\mbf x_0,\mbf x_0^{(i)})
\end{equation}
Particles can either be sampled directly from $p(\mbf x_0)$, in which case $w_0^{(i)} = 1/N \ \forall i$, or they can be sampled from an importance distribution $\pi(\mbf x_0)$, and set the weights as $w_0^{(i)} \propto p(\mbf x_0^{(i)}) / \pi(\mbf x_0^{(i)})$. It is worth noticing that in the particular case that the importance distribution is the uniform distribution, then $w_0^{(i)} \propto p(\mbf x_0^{(i)})$. The recursive formulation is obtained by applying Bayes' rule
\begin{align}
    p(\mbf X_k|&\mbf Y_k) = \dfrac{p(\mbf y_k| \mbf X_k,\mbf Y_{k-1})p(\mbf X_k| \mbf Y_{k-1})}{p(\mbf y_k|\mbf Y_{k-1})} \\
    =& \dfrac{p(\mbf y_k| \mbf X_k,\mbf Y_{k-1})p(\mbf X_k|\mbf X_{k-1}, \mbf Y_{k-1})p(\mbf X_{k-1}| \mbf Y_{k-1})}{p(\mbf y_k|\mbf Y_{k-1})} \label{reclaw}\\
    =& \dfrac{p(\mbf y_k, \mbf x_k |\mbf X_{k-1}, \mbf Y_{k-1})}{p(\mbf y_k|\mbf Y_{k-1})}     p(\mbf X_{k-1}|\mbf Y_{k-1})
\end{align}
It would be ideal to sample particles directly from the posterior distribution $p(\mbf x_k |\mbf X_{k-1}, \mbf Y_k)$, which is usually impossible (and it would mean that the problem is already solved). Therefore, having available samples and weights from the prior $p(\mbf X_{k-1} |\mbf Y_{k-1})$, we would get samples $\mbf x_k^{(i)}$ from a well-defined importance distribution 
$\pi(\mbf x_k |\mbf X_{k-1}, \mbf Y_k)$ and update the weights $w_k^{(i)}$ accordingly.

%%%%%%%%%%%%%%%%%%%%%%%%%%%%%%%%%%%%%%%%%%%%%%%%%%%%%%%%%%%%%%%%%%%%%%%%%%%%%%%%%%%%%%%%%
\subsection{Sequential Importance Sampling}
Sequential Importance Sampling (SIS) is a sequential Monte Carlo method. Recalling the assumption of working with a Markov process and independent random variables for the noises, Eq. \eqref{reclaw} leads to the proportionality relationship
\begin{equation}
    p(\mbf X_k|\mbf Y_k) \propto p(\mbf y_k| \mbf x_k)p(\mbf x_k|\mbf x_{k-1})p(\mbf X_{k-1}| \mbf Y_{k-1})
\end{equation}
This can be used to update the weights of the samples. First, we draw $\mbf X_{k-1}^{(i)}$ from the prior distribution, then we sample $\mbf x_k^{(i)}$ from the importance distribution $\pi(\mbf x_k |\mbf X_{k-1}, \mbf Y_k)$, such that $\mbf X_k^{(i)} = \left[\mbf X_{k-1}^{(i)} \quad \mbf x_k^{(i)} \right]$. Since the two parts are drawn independently, the importance distribution from which $\mbf X_k^{(i)}$ are drawn is the product between the distribution used to sample $\mbf x_k^{(i)}$ and the one used to sample $\mbf X_{k-1}^{(i)}$. Therefore, importance weights can be updated considering their proportionality to the ratio of the true distribution and the importance distribution:
\begin{align}
    w^{(i)}_k & \propto \dfrac{p(\mbf y_k| \mbf x_k^{(i)}) p(\mbf x_k^{(i)}|\mbf x_{k-1}^{(i)} ) p(\mbf X_{k-1}^{(i)}|\mbf Y_{k-1})}{\pi(\mbf x_k^{(i)}| \mbf X_{k-1}^{(i)},\mbf Y_k) \pi(\mbf X_{k-1}^{(i)}|\mbf Y_{k-1})} \\
    &= \dfrac{p(\mbf y_k| \mbf x_k^{(i)}) p(\mbf x_k^{(i)}|\mbf x_{k-1}^{(i)} ) }{\pi(\mbf x_k^{(i)}| \mbf X_{k-1}^{(i)},\mbf Y_k) } w_{k-1}^{(i)} \label{final}
\end{align}
This recurrence formulation shows how, given a prior set of $N$ points and weights relative step to $k-1$, the updated weights are calculated after drawing $N$ samples from the importance distribution $\pi(\mbf x_k^{(i)}| \mbf X_{k-1},\mbf Y_k)$. The posterior density function is, therefore, approximated as a PMF 
\begin{align}
    p(\mbf X_k | \mbf Y_k) = \sum^N_{i=1} w^{(i)}_k \delta(\mbf X_k, \mbf X_k^{(i)})
\end{align}
where marginalization is achieved by dropping the samples relative to old realizations of the state:
\begin{align}
    p(\mbf x_k | \mbf Y_k) = \sum^N_{i=1} w^{(i)}_k \delta(\mbf x_k, \mbf x_k^{(i)})
\end{align}
Various SIS algorithms differ in the selection of the importance distribution. The most common choice is also the simplest, where the prior PDF $p(\mbf x_k|\mbf x_{k-1})$ is selected, and the resulting methodology is the Bootstrap algorithm. The simplicity advantage of the method has a large inefficiency drawback, as the importance distribution completely ignores any information coming from the measurements $\mbf y_k$, such that many samples are drawn in regions where they are not needed. Obviously, the optimal solution would be to sample directly from the true posterior $p(\mbf x_k| \mbf X_{k-1}, \mbf Y_k)$, which is unknown and the aim of the estimation process. However, it is possible to use other filtering techniques to obtain an accurate approximation of the posterior and sample for it. Indeed, we could draw particles from posterior Gaussian distributions whose parameters have been calculated using the EKF approximation or the UKF solution (and so on) and have the SIS-EKF algorithm, the SIS-UKF, or even SIS-GSF.

%%%%%%%%%%%%%%%%%%%%%%%%%%%%%%%%%%%%%%%%%%%%%%%%%%%%%%%%%%%%%%%%%%%%%%%%%%%%%%%%%%%%%%%%%
%%%%%%%%%%%%%%%%%%%%%%%%%%%%%%%%%%%%%%%%%%%%%%%%%%%%%%%%%%%%%%%%%%%%%%%%%%%%%%%%%%%%%%%%%
%%%%%%%%%%%%%%%%%%%%%%%%%%%%%%%%%%%%%%%%%%%%%%%%%%%%%%%%%%%%%%%%%%%%%%%%%%%%%%%%%%%%%%%%%
\section{Filtering with Scouting Particles: the Scout Particle Filter} \label{SPFsec}
Following the SIS description, this paper offers a new approach to the selection of the importance distribution. Thanks to DA techniques, the measurement relation with the state of the system can be inverted to highlight the influence of the likelihood on the prior distribution. Knowing the covariance of the measurement noise, the main idea of the proposed technique is to delimit the region of the prior that could have produced the measurement outcome. This can be achieved by the creation of \textit{scout} particles, sampled directly from the measurement noise distribution in the measurement space and mapped back onto the state space. The scout particles give information regarding the location where the importance sampling distribution should draw particles with high density, improving both the filter's efficiency and accuracy.  

Given the system described in Eq. \eqref{system}, the Scout Particle Filter (SPF) is initialized with the state mean $\hat{\mbf x}_0$ and covariance $\mbf P_0$. Both the process noise covariance $\mathbb E [\boldsymbol \nu \boldsymbol \nu^T] = \mbf Q$ and the measurement noise covariance $\mathbb E [\boldsymbol \mu \boldsymbol \mu^T] = \mbf R$ are assumed known and independent between each other and from previous time steps. Whenever an observation becomes available, the SPF algorithm processes the measurement in the prediction and correction (or update) steps. The different parts of the filter described here are summed up in the software architecture layout, Figure \ref{figstr}, at the end of the chapter.

\subsection{The Prediction Step}
Given the current estimate $\hat{\mbf x}_k$ at step $k$, the prediction step propagates the state uncertainties forward in time. Following the Monte Carlo approach, the state PDF is approximated with $N_p$ particles that undergo the integration of the dynamics. This can be done efficiently in the DA framework \cite{jota}. The state polynomial is initialized considering as a variable the deviation from the mean:
\begin{equation}
    \mbf x_k(\delta \mbf x_k) = \hat{\mbf x}_k + \delta \mbf x_k
\end{equation}
that is, the state polynomial is centered at the current estimate.

Time propagation is applied directly to the polynomial according to the equations of motion:
\begin{equation}
    \mbf x_{k+1}(\delta \mbf x_k, \delta \boldsymbol \nu_k) = \mbf f (\mbf x_k(\delta \mbf x_k), \delta \boldsymbol \nu_k ) \label{prop}
\end{equation}
where $\delta \boldsymbol \nu_k$ indicates the influence of the process noise, expressed as a deviation vector. Since the noise is assumed with nil mean, the equality $\delta \boldsymbol \nu_k = \boldsymbol \nu_k $ holds, as the deviations have the same spread as drawing directly from the process noise distribution. The result of Eq. \eqref{prop} is a polynomial state transition map (PSTM) that connects state deviations at step $k$ to state realizations at step $k+1$. Therefore, state deviations can be drawn by the state PDF $p_{\mbf x_k}(\mbf x_k)$ centered around its mean. For example, if the state distribution is a Gaussian with given mean and covariance, $p_{\mbf x_k}(\mbf x_k) = \mathcal N(\hat{\mbf x}_k,\mbf P_k)$, deviations $\delta \mbf x_k^{(i)}$, with $i = 1,\dots,N_p$, can be drawn from $\mathcal N(\mbf 0,\mbf P_k)$. Likewise, $N_p$ process noise deviation vectors $\delta \boldsymbol \nu _k^{(i)}$ are drawn from the process noise distribution $p_{\boldsymbol \nu_k}(\delta \boldsymbol \nu_k)$. The Monte Carlo particle propagation typical of particle filters is here computed using polynomials, by evaluating the PSTM $N_p$ times. Therefore, the propagated set of particles is obtained as
\begin{equation}
    \mbf x_{k+1}^{(i)} = \mbf x_{k+1}(\delta \mbf x_k^{(i)}, \delta \boldsymbol \nu_k^{(i)}) \quad \forall i = 1,\dots,N_p
\end{equation}
where $N_p-1$ numerical propagations have been substituted by the computationally efficient polynomial evaluations. The computational advantages of the DA ensemble propagation are offered in \cite{jota}. The predicted mean and covariance of the propagated distribution are calculated as 
\begin{align}
    \hat{\mbf x}_{k+1}^{-} &= \sum_{i=1}^{N_p} w_{k+1}^{(i)} \mbf x_{k+1}^{(i)} \\
    \mbf P_{k+1}^{-} &= \sum_{i=1}^{N_p} w_{k+1}^{(i)} (\mbf x_{k+1}^{(i)} - \hat{\mbf x}_{k+1}^{-})(\mbf x_{k+1}^{(i)} - \hat{\mbf x}_{k+1}^{-})^T
\end{align}
where $w_{k+1}^{(i)} = 1/N_p$, as particles are assumed to be drawn directly from the state distribution. The same results would have been achieved by uniform sampling in the space domain and assigning its probability as the weight to each particle.

%%%%%%%%%%%%%%%%%%%%%%%%%%%%%%%%%%%%%%%%%%%%%%%%%%%%%%%%%%%%%%%%%%%%%%%%%%%%%%%%%%%%%%%%%
\subsection{The Correction Step}
The particles have been propagated up to a time when an observation $\tilde{\mbf y}_{k+1} $ becomes available from the sensor, and the SPF performs the measurement update step to correct its own prediction. Indeed, a new state polynomial is initialized around the predicted mean:
\begin{equation}
    \mbf x_{k+1}(\delta \mbf x_{k+1}) = \hat{\mbf x}_{k+1}^{-} + \delta \mbf x_{k+1}
\end{equation}
Following the theory presented in Eq. \eqref{mapex}, the measurement polynomial is evaluated by applying the nonlinear measurement equations directly onto the state polynomials:
\begin{align}
    \mbf y_{k+1} (\delta \mbf x_{k+1}) &= \mbf h_{k+1} (\mbf x_{k+1}(\delta \mbf x_{k+1}) ) \\ 
    &= \bar{\mbf y}_{k+1} + \mathcal M_{\hat{\mbf x}_{k+1}^{-}}( \delta \mbf x_{k+1}) \label{mapmeas}
\end{align}
where the map $\mathcal M_{\hat{\mbf x}_{k+1}^{-}}( \delta \mbf x_{k+1})$, centered in $\hat{\mbf x}_{k+1}^{-}$, connects deviations vectors from the state space onto the measurement space. This polynomial is made of a composition of Taylor polynomials up to a certain truncation order. 

The main idea of the SPF is to randomly sample the measurement space, knowing the noise covariance level, and to map this information back to the state prior distribution in order to have an efficient drawing during SIS. Regardless of the generation of particles, which will be later addressed, the measurement map needs to be inverted to connected deviations $\delta \mbf y_{k+1}$ to the state deviation $\delta \mbf x_{k+1}$ that generates them: 
\begin{equation}
    \mathcal W_{\bar{\mbf y}_{k+1}}(\delta \mbf y_{k+1}) = \big( \mathcal M_{\hat{\mbf x}_{k+1}^{-}}( \delta \mbf x_{k+1}) \big)^{-1}
\end{equation}
where $\mathcal W_{\bar{\mbf y}_{k+1}}(\delta \mbf y_{k+1})$ indicates the inverted polynomial map that maps the state space from the measurement space. The map inversion can be computed directly on the polynomials and can be achieved efficiently in a computer environment using DA techniques \cite{berz}. However, for the map inversion to be successful, the measurements need to be independent of each other, and the equality $n = m$ needs to hold; that is, there is an equal number of measurements and states. Moreover, it is assumed that no singularities break the inversion process. These assumptions can be connected to the requirements for matrix inversions, where, to be invertible, any matrix must be square and full rank. As shown in the following section, the harsh limitation on the required number of measurements can be overcome. 

The measurement $\tilde{\mbf y}_{k+1} $ is received from the sensors. Considering the assumption of additive noise, the true measurement, i.e., the numerical outcome that is computed by applying the measurement equations to the true state of the system, lies in the vicinity according to the measurement noise PDF. Therefore, by drawing particles according to the noise distribution centered in $\tilde{\mbf y}_{k+1}$, the algorithm delimits a region in the measurement space that includes the true measurement. Scout particles are drawn according to the measurement noise distribution $p_{\boldsymbol \mu_{k+1}}(\boldsymbol \mu_{k+1})$ translated to be centered at $\tilde{\mbf y}_{k+1} $. It is worth noticing that no assumption on the type of distribution has been made, as the algorithm handles any colored noise. In the particular case of zero-mean Gaussian noise, scout particles are drawn from $\mathcal N(\tilde{\mbf y}_{k+1},\mbf R_{k+1})$. Let us call with $\mbf y^{(i)}_s$ the scout particles: their deviation vectors from the center can be evaluated as
\begin{equation}
    \delta \mbf y_s^{(i)} = \mbf y_s^{(i)} - \bar{\mbf y}_{k+1} \quad \forall i= 1,\dots,N_s \label{dyk}
\end{equation}
where $N_s$ is the number of scout particles, selected at each time step. By drawing scout particles directly from the measurement noise distribution, they all share the same weight. The state deviation vector that would have generated the scout particle in the measurement space is calculated through polynomial evaluation of the inverted map:
\begin{equation}
    \delta \mbf x_s^{(i)} = \mathcal W_{\bar{\mbf y}_{k+1}}(\delta \mbf y_s^{(i)}) \quad \forall i= 1,\dots,N_s
\end{equation}
Therefore, the set of scout particles in the state space is computed by summing the predicted mean to the deviations
\begin{equation}
    \mbf x_s^{(i)} = \hat{\mbf x}_{k+1}^{-} + \delta \mbf x_s^{(i)}  \quad \forall i= 1,\dots,N_s
\end{equation}
The distribution of the set of scout particles encapsulates the sector of the prior that could have generated the measurement, thus delimiting the region that is most likely to have the true state. Therefore, the location of the scout particles can be used to obtain a well-chosen importance distribution to apply the SIS update. First, the Scout mean and covariance matrix are evaluated
\begin{align}
    \hat{\mbf x}_s &= \sum_{i=1}^{N_s} \mbf x_s^{(i)} \\
    \mbf P_s &= \sum_{i=1}^{N_s} (\mbf x_s^{(i)} - \hat{\mbf x}_s)(\mbf x_s^{(i)} - \hat{\mbf x}_s)^T
\end{align}
where all the scout particles share the same weight $1/N_s$ as they have been drawn directly from the noise PDF. The scout mean tells where the likelihood distribution meets the state prior, while the scout covariance gives information regarding the spread of the posterior PDF. Thus, a well-informed importance distribution for the SIS algorithm can be chosen directly from the scout variables, and there are two main options. The first option is to sample from a uniform distribution centered at $\hat{\mbf x}_s$ over the $3\sigma_s$ standard deviation support dictated by $\mbf P_s$. That is, particles are drawn from the $n$-dimensional parallelepiped centered at the scout mean with sides $\trace(6\mbf S_s)$, where $\mbf S_s \mbf S_s^T = \mbf P_s$. The second option is to sample state particles from the Gaussian distribution fitted from the scout particles. Thus, given the particle $\mbf x^{(i)}_{k+1}$ sampled according either distribution, its importance weight is
\begin{align}
    w^{(i)}_{SIS} &= \text{unif}\big(\mbf x^{(i)}_{k+1}|\hat{\mbf x}_s - 3\trace(\mbf S_s), \hat{\mbf x}_s + 3\trace(\mbf S_s)\big) \label{sis1}\\
    w^{(i)}_{SIS} &= \mathcal N(\mbf x^{(i)}_{k+1}|\hat{\mbf x}_s,\mbf P_s)\label{sis2}
\end{align}
which represents the probability of the $i$th particle according to the selected importance distribution. It is worth noticing that for the uniform distribution case, after normalization, all the particles share the same weight of $1/N_u$, being $N_u$ the total number of samples used for the update. 

It is worth noticing that the uniform and Gaussian selections are the most computationally efficient, and they achieve their goal of binding the region of the prior that generates the posterior. However, if requested or necessary for a specific application, it is possible to fit the importance distribution via least square or divide it into multiple Gaussians via Expectation Maximization according to the Gaussian Multiple Model mathematics.  

Particles have been created according to the importance distribution. The particle image in the measurement codomain is evaluated efficiently in the DA framework using deviations
\begin{align}
    \delta \mbf x_{k+1}^{(i)} &= \mbf x^{(i)}_{k+1} - \hat{\mbf x}_{k+1}^{-} \quad \forall i = 1,\dots,N_u
\end{align}
and polynomial evaluations
\begin{align}
    \mbf y_{k+1}^{(i)} &= \mbf y_{k+1} (\delta \mbf x_{k+1}^{(i)}) \quad \forall i = 1,\dots,N_u
\end{align}
The likelihood weight of the particle is computed considering its probability given the observation and the measurement noise distribution:
\begin{equation}
    w^{(i)}_{LIK} = p_{\boldsymbol \mu_{k+1}}(\mbf y_{k+1}^{(i)} - \tilde{\mbf y}_{k+1}) \quad \forall i = 1,\dots,N_u
\end{equation}
For clarity, in the particular case of null mean Gaussian measurement noise with given covariance matrix $\mbf R_{k+1}$, the likelihood weight is computed as
\begin{equation}
    w^{(i)}_{LIK} = \exp{ \Big(-\dfrac{1}{2} (\mbf y_{k+1}^{(i)} - \tilde{\mbf y}_{k+1})^T \mbf R^{-1}_{k+1} (\mbf y_{k+1}^{(i)} - \tilde{\mbf y}_{k+1})} \Big)
\end{equation}
as the normalizing factor of the PDF, being constant for each weight, can be omitted. 

Finally, the updated weight of the particles can be evaluated by merging the likelihood influence with the predicted distribution. Assuming a Gaussian prior, the initial prior weight of each sample is
\begin{equation}
     w^{(i)}_{PRE} = \mathcal N(\mbf x^{(i)}_{k+1}|\hat{\mbf x}_{k+1}^{-}, \mbf P_{k+1}^{-})   \quad \forall i = 1,\dots,N_u
\end{equation}
such that it can corrected using ratios of probabilities according to the SIS formulation:
\begin{equation}
     w^{(i)}_{POST} = \dfrac{w^{(i)}_{LIK}w^{(i)}_{PRE}}{w^{(i)}_{SIS}}   \quad \forall i = 1,\dots,N_u
\end{equation}
The $N_u$ weights are then normalized such that they sum to unity. The Appendix shows a convenient way that avoid numerical issues and work out normalization when some particles are located in low-density regions. By comparing this final relation for the update of the weights with Eq. \eqref{final}, it is worth noticing how each contribution evaluated separately is merged together following Bayes' rule. Moreover, the SPF weight update leaves freedom to include any methodology for the change of particle weight during the prediction, as the two steps are decoupled. An accurate prediction algorithm from the literature can be merged into the SPF without modifying the overall scouting methodology. 

Lastly, the state estimate and accuracy confidence level of the current step is given as the weighted mean of the SIS particles:
\begin{align}
    \hat{\mbf x}_{k+1}^{+} &= \sum_{i=1}^{N_u} w_{POST}^{(i)} \mbf x_{k+1}^{(i)} \\
    \mbf P_{k+1}^{+} &= \sum_{i=1}^{N_u} w_{POST}^{(i)} (\mbf x_{k+1}^{(i)} - \hat{\mbf x}_{k+1}^{+})(\mbf x_{k+1}^{(i)} - \hat{\mbf x}_{k+1}^{+})^T
\end{align}
The correction step is concluded and the observations from the sensor have been merged into the filter's prediction. Before propagating to the following step and elaborating a new measurement, the SPF can adopt, if necessary, any resampling algorithm to overcome impoverishment issues. As a default technique, the SPF chooses the resampling step of the GPF \cite{gpf}. This resampling step is easy to implement and it is performed by drawing from a Gaussian: this assures particle diversity, but loses information regarding the non-Gaussian shape of the posterior distribution resulting from the nonlinear dynamics. Nevertheless, if needed, a more conserving resampling technique can be adopted without influencing the novelties proposed by the SPF.

%%%%%%%%%%%%%%%%%%%%%%%%%%%%%%%%%%%%%%%%%%%%%%%%%%%%%%%%%%%%%%%%%%%%%%%%%%%%%%%%%%%%%%%%%
\subsection{The Scout Particle Filter for Non-Square Applications}
The creation of the inverted polynomial map used by the scout particles to define the region of the prior where the importance distribution should have high density requires an equal amount of measurements and states, $m = n$. This assumption is limiting the number of applications where the SPF can be applied. In the case that there are more measurements than states, $m>n$, the proposed technique works fine by selecting the measurements up to the number of states to invert the map, as long as they are independent of each other. Measurements are picked to maximize the observability of the map, resulting in increased stability and accuracy \cite{obs}. After scouting, the SPF applies its update using the whole measurement set. The most challenging scenario is when there are fewer measurements than states, case $m<n$. Previous works resolved the issue by storing measurements until the required amount is reached \cite{rev}. In doing so, the inverted mapping maps together multiple time steps, and it relies on the assumption that the number of states is a multiple of the dimension of the measurement. A different solution is offered here that does not require the filter to wait, such that the SPF can perform the update step whenever an observation becomes available. 

Given independent measurements, a non-invertible polynomial map happens when $m<n$. In order to counter this issue, the SPF forces the map to be square by creating fake observations. Therefore, an auxiliary function $\mbf q_{k+1}(\mbf x_{k+1})$ is selected to fill in for the missing measurements, such that the augmented polynomial map
\begin{align}
    \mbf y_{aug} (\delta \mbf x_{k+1}) &= 
    \begin{bmatrix}
        \mbf h_{k+1} (\mbf x_{k+1}(\delta \mbf x_{k+1}) ) \\
        \mbf q_{k+1} (\mbf x_{k+1}(\delta \mbf x_{k+1}) ) 
    \end{bmatrix}\\
    &= \bar{\mbf y}_{aug} + \mathcal M_{\hat{\mbf x}_{k+1}^{-}}( \delta \mbf x_{k+1}) 
\end{align}
is square. This formulation replaces Eq. \eqref{mapmeas}, assuring the definition of an invertible map. Indeed, even the center of the measurement polynomial has been augmented to match the new dimensions, where it is composed by two separate parts:
\begin{align}
    \bar{\mbf y}_{aug} &= 
    \begin{bmatrix}
        \bar{\mbf y}_{k+1} \\
        \mbf q_{k+1} (\hat{\mbf x}_{k+1}^{-}) 
    \end{bmatrix}
    = 
    \begin{bmatrix}
        \bar{\mbf y}_{k+1} \\
        \bar{\mbf q}_{k+1}
    \end{bmatrix}
\end{align}
In the square case, the scout points $\mbf y_{k+1}^{(i)}$ were drawn from the measurement noise distribution centered at the given observation. The generation of scout particles can be slightly modified to include the fictitious measurements. The first part of the vector (the first $m$ components) of a scout particle is sampled using $p_{\boldsymbol \mu_{k+1}}(\boldsymbol \mu_{k+1})$, likewise for the square map. The second part of the vector (the remaining $n-m$ components) is instead sampled by the Gaussian distribution $\mathcal N(\bar{\mbf q}_{k+1},\mbf P_q)$, where 
\begin{align}
    \mbf H_{k+1} &= \dfrac{\partial \mbf q_{k+1}}{\partial \mbf x_{k+1}}\bigg\rvert_{\hat{\mbf x}_{k+1}^{-}} \\
    \mbf P_q &= \mbf H_{k+1} \mbf P_{k+1}^{-} \mbf H_{k+1}^T \label{pekf}
\end{align}
Basically, using the Jacobian $\mbf H_{k+1}$ of the fictitious measurement function, the spread level of the fake distribution is approximated using the linearized covariance propagation commonly used in the EKF. Equation \eqref{pekf} is one of the simplest and most computationally efficient ways to evaluate the transformed covariance. However, if necessary, due to the selection of a highly nonlinear arbitrary function $\mbf q_{k+1}(\mbf x_{k+1})$, Eq. \eqref{pekf} can be replaced by the more accurate uncertainty transformation. Nevertheless, mere linearization should suffice, as the best choice for $\mbf q_{k+1}(\mbf x_{k+1})$ is the identity function, where the fake observation directly measures the state of the system and $\mbf H _{k+1}$ is composed of an identity matrix. Conceptually, the $N_s$ scouting particles have been drawn considering the actual measurement noise and a fake Gaussian noise with covariance predicted by the filter. Therefore, for the non-square case, Eq. \eqref{dyk} is rewritten as
\begin{equation}
    \delta \mbf y_s^{(i)} = \mbf y_s^{(i)} - \bar{\mbf y}_{aug} \quad \forall i= 1,\dots,N_s 
\end{equation}
and the remaining part of the SPF filtering algorithm continues unaltered. 

The creation of fictitious measurements is a clever way to achieve inversion when the system does not receive the supposed amount of information. The fake observation coming from the sensor could either be the center or the mean of fictitious measurement PDF, as the performance of the filter is very lightly affected by this decision due to the large value of $\mbf P_q$ when compared to $\mbf R_{k+1}$. Indeed, the added measurement function aims to help the scout particles travel to the prior distribution, not to improve the likelihood estimation.

%%%%%%%%%%%%%%%%%%%%%%%%%%%%%%%%%%%%%%%%%%%%%%%%%%%%%%%%%%%%%%%%%%%%%%%%%%%%%%%%%%%%%%%%%
\subsection{Software Architecture}
\begin{figure*}[t]
    \centering
    \includegraphics[width=\textwidth]{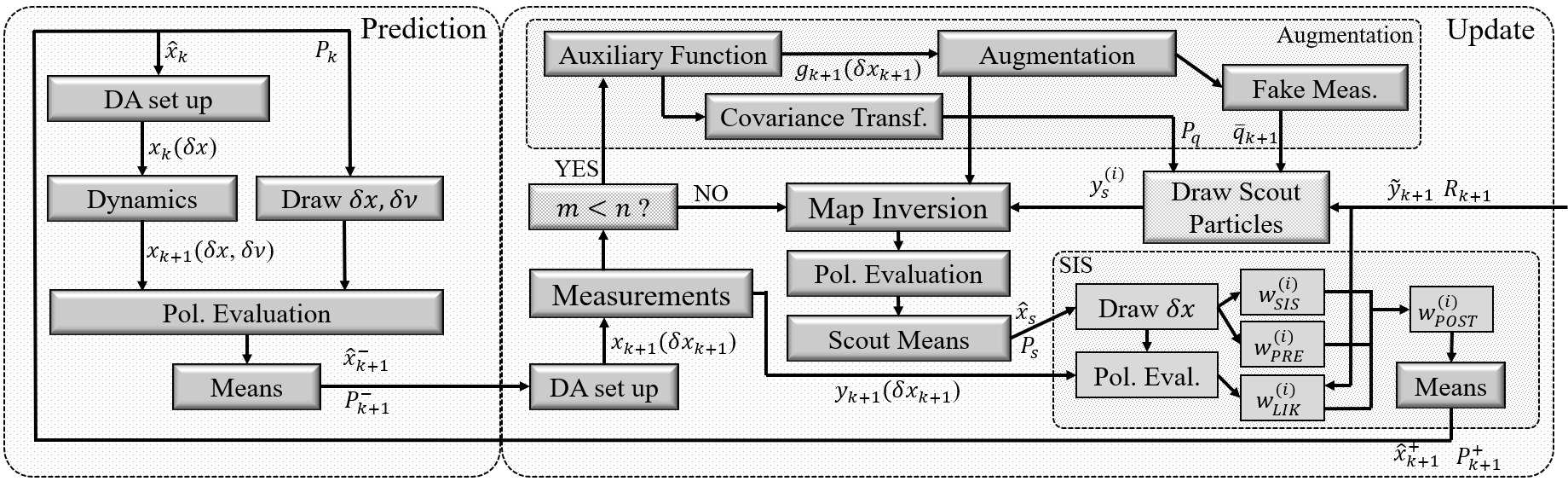}
    \caption{Scout Particle Filter Software Architecture.}
    \label{figstr}
\end{figure*}
The overall structure of the Scout Particle Filter is outlined in Fig. \ref{figstr}, where the filtering algorithm has been deconstructed to show connections and components. The primary separation of the algorithm is the division between the prediction and the update step. The two steps are connected merely by the exchange of the current predicted mean and covariance. As previously described, the update has a significant improvement with respect to the literature. Indeed, the DA mathematics supports the SIS algorithm, preceded by the creation of the polynomial maps and their inversions. The only information coming from the external world is the observation from the sensor with relative noise covariance level. This measurement is used to improve the accuracy of the estimated state PDF.  

As the figure shows, particles are drawn from distributions on three different occasions during one single iteration of the filtering algorithm. Therefore, the tuning of the algorithm is performed by selecting the number of particles during prediction, $N_p$, for the update, $N_u$, and scout particles, $N_s$. 

The software architecture underlines the case where measurement augmentation is required. In the case of a few measurements, $m<n$, the SPF enters the augmentation part where fictitious observations are created such that the polynomial map can be inverted and scout particles can travel to the state space.

After undergoing the inverted map through polynomial evaluation, the scout particles delimit the region of the prior that could have generated the measurement. Therefore, it is obvious that the benefits of scouting are more beneficial whenever there is a poor knowledge of the prior state distribution. If the filter has given a very accurate predicted PDF, the scout particles will delimit the whole state distribution as a high-density region. Thus, scouting helps the filter, especially during the initial steps and transient period of the estimation process, before reaching convergence. Indeed, when convergence is reached, the SPF can decide to stop generating scouting particles and apply any common particle filter update, reducing the number of computations. In a similar manner, the SPF is particularly effective when the system is affected by heavy process noises or when the subsequent measurement acquisition is separated by long propagation times, factors that tend to enlarge the spread of the state PDF affected by the nonlinear dynamics. 

The derived filtering technique proposes an importance sampling algorithm informed by the measurement noise. The implementation in the DA framework opens the SPF algorithm to improvement by merging with other DA techniques already validated in the literature, such as Automatic Domain Splitting, where the convergence and accuracy of the Taylor polynomial expansion are checked at the edges of the domain and split in half whenever the approximation error is larger than a given threshold. Merging ADS with the SPF improves the prediction step of the filter, especially for vast initial uncertainties and extremely long propagation times \cite{losacco}. Such derivation is left for future developments as it is currently out of the scope of the paper.

%%%%%%%%%%%%%%%%%%%%%%%%%%%%%%%%%%%%%%%%%%%%%%%%%%%%%%%%%%%%%%%%%%%%%%%%%%%%%%%%%%%%%%%%%
%%%%%%%%%%%%%%%%%%%%%%%%%%%%%%%%%%%%%%%%%%%%%%%%%%%%%%%%%%%%%%%%%%%%%%%%%%%%%%%%%%%%%%%%%
%%%%%%%%%%%%%%%%%%%%%%%%%%%%%%%%%%%%%%%%%%%%%%%%%%%%%%%%%%%%%%%%%%%%%%%%%%%%%%%%%%%%%%%%%
\section{Numerical Applications} \label{appsec}
Five different numerical examples are shown here to demonstrate the benefits of the Scout Particle Filter. The SPF is compared to other particle filtering, such as the BPF, and other importance sampling filters, like the SIS-EKF and the SIS-UKF. Generally, the SIS version of the common EKF and UKF is more accurate than the classic EKF and UKF algorithm, as importance sampling adds an additional step to improve accuracy from the updated distribution of the filters, where the Gaussian approximation of the posterior is used as importance distribution \cite{p4, rev2}. 

When comparing the performance of particle filters, it is critical to evaluate them both in terms of accuracy and efficiency. Since these filters are based on the sequential Monte Carlo method, it is natural to achieve higher accuracy levels by increasing the number of particles drawn from the importance distribution. Therefore, the SPF will also be compared in terms of efficiency by calculating the percentage of efficient particles over the whole ensemble. The best importance distributions have high density in the same regions as the true posterior, but, inevitability, the same samples are drawn from locations with almost null probability of belonging to such PDF. An index that evaluates the number of effective particles is
\begin{equation}
    N_{eff} = \dfrac{1}{\sum^{N_u}_{i=1} (w_{POST}^{(i)})^2}
\end{equation}
from \cite{neff}, which is an estimate of the measure of the degeneracy of the algorithm presented in \cite{neff2} as the ``rule of thumb" for importance sampling
\begin{equation}
    \tilde{N}_{eff} = \dfrac{N_u}{1+\text{var}_{\pi(\mbf x)}(w(\mbf x))} = \dfrac{N_u}{\mathbb E_{\pi(\mbf x)}[w(\mbf x)^2]} \leq N_u
\end{equation}
that defines the effective sample size. Alternatives for effective sample size (ESS) measures are described in \cite{neff3}. If all particles have the same weight $1/N_u$, then the equivalency $N_{eff} = N_u $ holds. However, SIS methods tend to suffer from impoverishment and degeneracy where a few samples gather most of the weight, and $N_{eff} $ tends to one. Therefore, most Particle Filters are Sequential Importance Sampling with Resampling (SISR) algorithms, as they resample the particles whenever $N_{eff} $ falls below a defined threshold. For example, the BPF usually resamples at each step, as its importance distribution is poorly chosen. Consequently, the numerical applications here reported compare the efficiency of each filter in terms of the percentage of efficient particles:
\begin{equation}
  \Psi  = 100\dfrac{N_{eff}}{N_u}\%
\end{equation}
evaluated considering the ratio of useful samples over the total ensemble drawn.

%%%%%%%%%%%%%%%%%%%%%%%%%%%%%%%%%%%%%%%%%%%%%%%%%%%%%%%%%%%%%%%%%%%%%%%%%%%%%%%%%%%%%%%%%
\subsection{Range-Angle Measurement}
\begin{figure*}
    \centering
    \includegraphics[width=\textwidth]{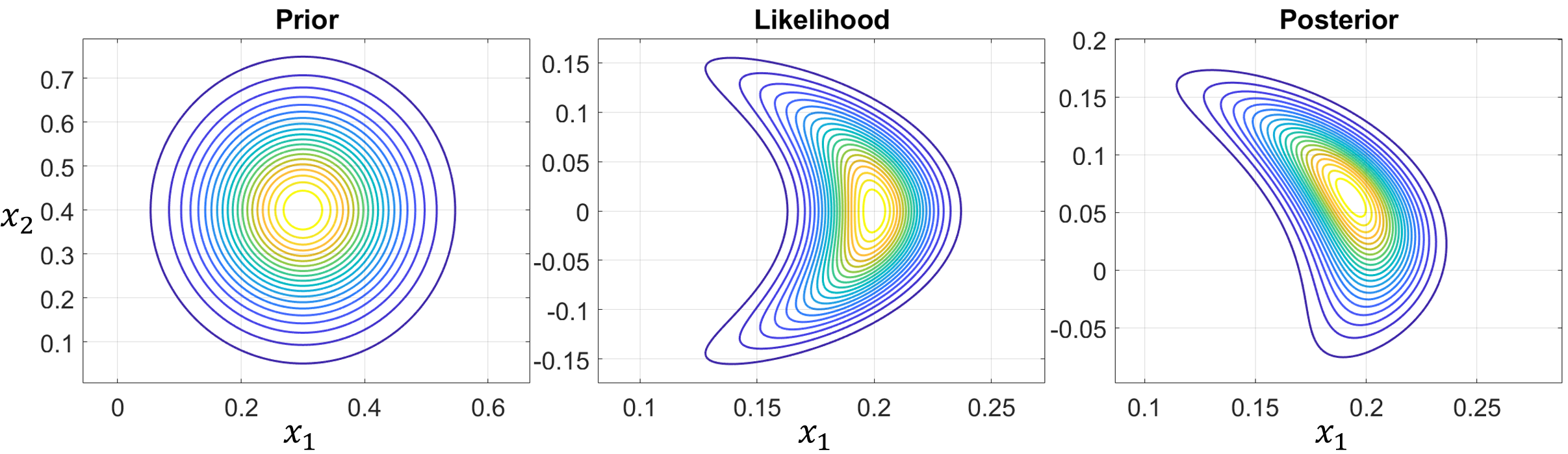}
    \caption{Prior, likelihood, and true posterior probability density functions.}
    \label{figpdfs}
\end{figure*}

The first scenario is a two-dimensional static application aimed to show the benefits of the scout particles. Consider a range and angle measurement problem where the prior is a Gaussian PDF:
\begin{equation}
    \mbf x = \mathcal N \Bigg( \mbf x; 
    \begin{bmatrix}
        0.3 \\ 0.4
    \end{bmatrix},
    \begin{bmatrix}
        0.01 & 0 \\ 0 & 0.02
    \end{bmatrix}
    \Bigg)
\end{equation}
and the sensors provide measurements from the origin
\begin{align}
    y_1 &= ||\mbf x|| + \mu_1 \label{range} \\ 
    y_2 &= \arctan \Big(\dfrac{x_2}{x_1}\Big)  + \mu_2 \label{angle}
\end{align}
where $\boldsymbol \mu$ is zero mean Gaussian noise. The range noise has a standard deviation of $\boldsymbol \sigma_{11} = 0.015$, while the bearing angle standard deviation is $\boldsymbol \sigma_{22} = 20$ degrees. Figure \ref{figpdfs} gives a visual representation of the state prior and true posterior distribution, as well as the likelihood, given as observation the numerical values of $\tilde{\mbf y} = [0.2 \ \ 0]^T $.

\begin{figure*}
    \centering
    \includegraphics[width=\textwidth]{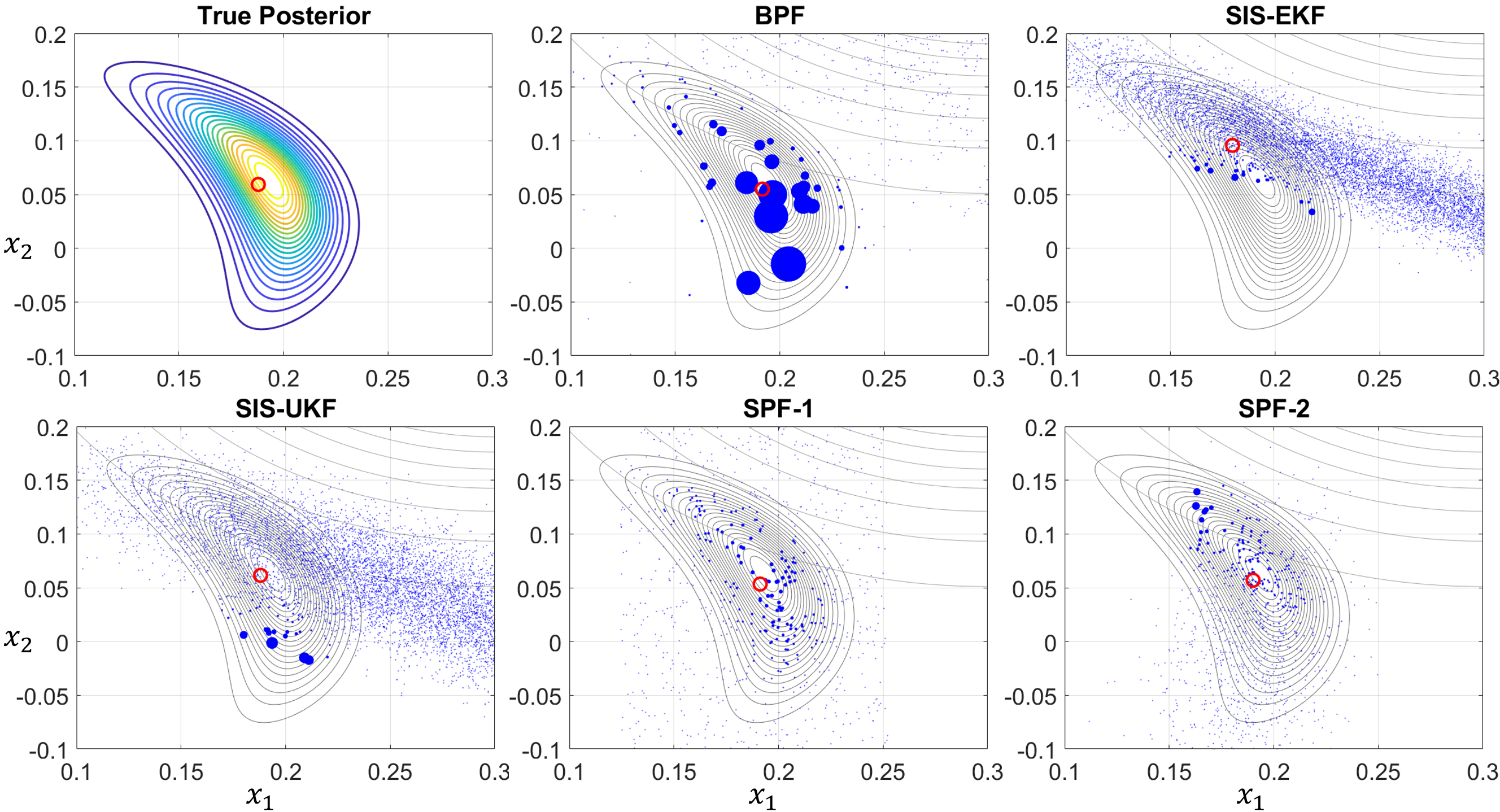}
    \caption{Different particle filter approximation of the true posterior distribution.}
    \label{visual}
\end{figure*}

Different particle filters based on SIS have been applied to the problem, and their performance has been compared in terms of accuracy and efficiency. Figure \ref{visual} shows the different weights that each particle receives after update by modifying the particle size: indeed, a bigger point corresponds to a particle with a larger weight. The figure reports the state space only around the true posterior distribution, reported in light gray in the background to help facilitate visual comparisons. The estimate of each filter, calculated as a weighted mean, is displayed as a red circle. The BPF has been implemented using $10^4$ particles: only very few have enough weight to be visualized in the figure. Most of them are outside the posterior distribution domain, as they are sampled centered from the prior mean. The BPF graph is a perfect example of particle impoverishment, where only a handful of particles (0.21\% in the presented case) positively contribute to the state estimation. A particle filter that does not suffer impoverishment has the size of its particle close together, without extremes. The SIS-EKF and SIS-UKF are the subsequent steps to improve estimation: the importance distribution is the updated Gaussian PDF calculated by the EKF and UKF. However, even if they show better particle efficiency than the BPF, many particles are drawn in low-density zones outside the posterior domain. In order to achieve valuable results, the SIS-EKF and SIS-UKF have been computed with $10^4$ particles. Lastly the Scout Particle Filter has been run in its two different variations, i.e., considering a uniform distribution from the scout particles in SPF-1, following Eq. \eqref{sis1}, or a Gaussian PDF in SPF-2, according to Eq. \eqref{sis2}. For both cases, only $N_s = 100$ scout particles have been drawn to scout the likelihood, such that a total of $N_u = 10^3$ samples is enough to achieve an accurate update. Looking at the figure, the SPF particles are efficiently drawn near the true posterior, so they share weight accordingly. In particular, using the uniform distribution, SPF-1 particles get heavier the closer they are to the true posterior peak, perfectly representing the distribution as a PMF.

\begin{figure}[ht]
    \centering
    \includegraphics[width=0.48\textwidth]{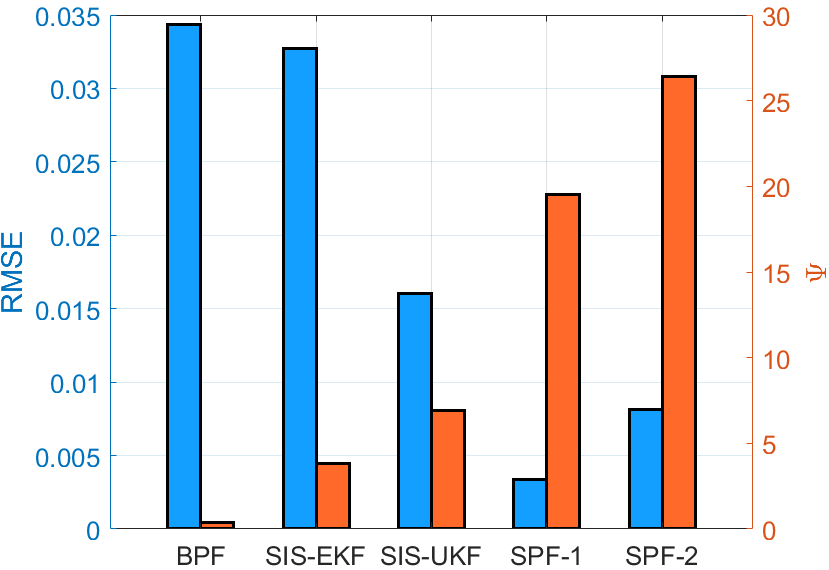}
    \caption{RMSE and efficiency comparison among particle filters.}
    \label{rmse1}
\end{figure}

A more accurate performance comparison is pursued by completing a Monte Carlo analysis and computing each filter's root mean square error (RMSE) and the relative $\Psi$ factor. Therefore, the measurement update, with constant parameters, has been repeated for each filter $N_{MC} = 10^5$ times, and the RMSE is computed as
\begin{equation}
    RMSE = \sqrt{\dfrac{1}{N_{MC}} \sum_{i=1}^{N_{MC}} (\hat{\mbf x}_i - \mbf x_{T})^T(\hat{\mbf x}_i - \mbf x_{T})  }
\end{equation}
where $\hat{\mbf x}_i$ is the estimate of the current Monte Carlo run and $\mbf x_{T}$ is the mean of the true posterior distribution. For a fair comparison, all filters have assigned the same number of particles, 1000. The SPF uses 50 particles in the scouting. Figure \ref{rmse1} is a histogram of the error level of the filters, blue bars. In the same graph, the efficiency of particle sampling is reported using $\Psi$, which is an indicator of the ratio of effective particles. Thus, the $\Psi$ factor is a measure of how well the filtering algorithm has chosen the importance distribution. As expected, the BPF has terrible efficiency and poor accuracy. Introducing the SIS algorithm, using the EKF and UKF, improves the performance of the filters, but only slightly due to bad estimates from the linear estimators. The figure shows that the performance of the SPF is superior with respect to the other filters, both in terms of accuracy and efficiency. Between the scouting filters, SPF-1 achieves more accurate estimates than its counterpart; however, SPF-2 proves that Gaussian sampling around the scout mean is more efficient in generating samples, meaning that it could achieve similar accuracy levels with fewer particles. 

%%%%%%%%%%%%%%%%%%%%%%%%%%%%%%%%%%%%%%%%%%%%%%%%%%%%%%%%%%%%%%%%%%%%%%%%%%%%%%%%%%%%%%%%%
\subsection{Range Measurement}
\begin{figure*}
    \centering
    \includegraphics[width=\textwidth]{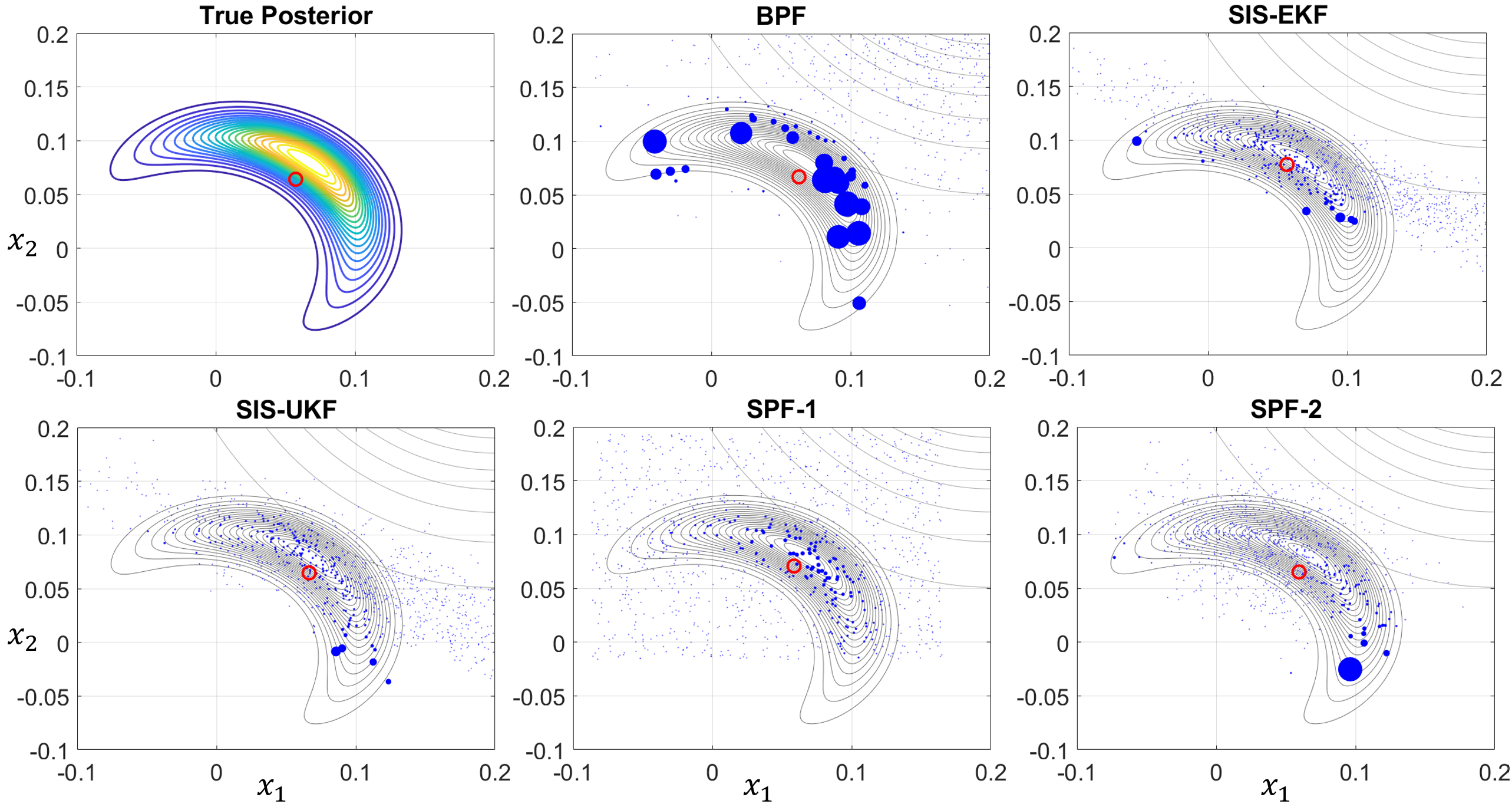}
    \caption{Different particle filter approximation of the true posterior distribution: range only.}
    \label{visual_range}
\end{figure*}

The application is now repeated after removing one of the measurements, considering a Gaussian state prior  
\begin{equation}
    \mbf x = \mathcal N \Bigg( \mbf x; 
    \begin{bmatrix}
        0.2 \\ 0.4
    \end{bmatrix},
    \begin{bmatrix}
        0.01 & 0 \\ 0 & 0.02
    \end{bmatrix}
    \Bigg)
\end{equation}
Indeed, the filters perform their update considering the singular scalar range observation, which has been selected as $\tilde y = 0.1$, calculated according to Eq. \eqref{range}, with noise standard deviation $\sigma_R = 0.015$. Figure \ref{visual_range} shows the results of the estimators for the new application. Lacking angle information, the true posterior distribution is more curved than the previous case, which penalizes linear estimators such as the EKF and the UKF. Thus, the two SIS particle filters that rely on such Gaussian assumption have poor performance as the Gaussian mean is far from the mean of the true posterior, red circle of the first graph. The BPF is, once again, poorly sampled as the observation has moved the likelihood on the edges of the prior distribution. Lastly, the two SPF filters show the best behavior. They create a fictitious angle measurement to invert the polynomial map, such that the scout particles are able to undergo polynomial evaluation. In particular, SPF-1 scans the whole posterior domain, as the uniform distribution creates a rectangle from which particles are drawn. On the contrary, SPF-2 concentrates its particles closer to the mean of the true posterior thanks to the Gaussian approximation of the scout samples. While the identity measurement could have been chosen as the fictitious measurement, the selection of the angle completes the polar representation of the state. It is connected to higher observability of the system, making the inverted map more robust and accurate. 

\begin{figure}[ht]
    \centering
    \includegraphics[width=0.48\textwidth]{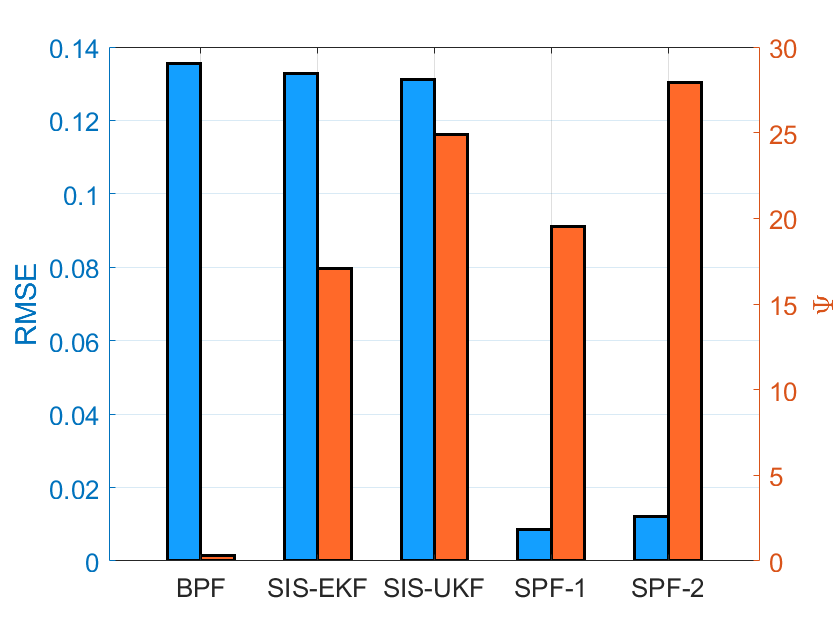}
    \caption{RMSE and efficiency comparison among particle filters: range only}
    \label{rmse2}
\end{figure}

The performance of the filters is compared in Fig. \ref{rmse2} with a $10^5$ Monte Carlo analysis. The figure shows that the SPF estimators are much more accurate than the other particle filters, more than an entire order of magnitude. The better estimation of the posterior for samplings lowers the RMSE. In fact, the SIS-UKF is more accurate than the SIS-EKF, which, in turn, is more precise than the BPF. However, their accuracy is far worse than the scout filters.

Moreover, Fig. \ref{rmse2} shows an interesting behavior between accuracy and efficiency. The SPF-1 is about 27.5\% more accurate than SPF-2, which is more efficient with its particle generation, as the $\Psi$ factor comparison proves. Therefore, a trade-off between the desired level of accuracy and the number of particles, which corresponds to the computational burden, can be performed when selecting the importance distribution of the SPF. Lastly, the figure underlines that there is no correlation between efficiency and accuracy: the SIS-UKF has a higher $\Psi$ than the SPF-1, but its RMSE level is much larger. 

%%%%%%%%%%%%%%%%%%%%%%%%%%%%%%%%%%%%%%%%%%%%%%%%%%%%%%%%%%%%%%%%%%%%%%%%%%%%%%%%%%%%%%%%%
\subsection{Projectile Tracking}
A time-dependent application is offered here. The SPF algorithm is applied to track the trajectory of a projectile launched with a given velocity. The system is affected by strong winds that randomly push the projectile. This application aims to show the benefits of scouting on systems affected by high-level process noises, which inflate the state distribution at each propagation between two separate measurement acquisitions. Thus, the advantages of scouting can be appreciated even for linear dynamics. Consider the system modeled by the equations of motion 
\begin{align}
    \mbf x_{k+1} = 
    \begin{bmatrix}
        1 & 0 & \Delta t & 0 \\
        0 & 1 & 0 & \Delta t \\
        0 & 0 & 1 & 0 \\
        0 & 0 & 0 & 1 \\
    \end{bmatrix} \mbf x_k +
    \begin{bmatrix}
        0  \\
        -\dfrac{1}{2}g\Delta t^2 \\
        0 \\
        -g \Delta t \\
    \end{bmatrix}
    + \boldsymbol \nu_k
\end{align}
where $g$ represents the gravity constant and, with $\Delta t = 0.2$ seconds,  $\boldsymbol \nu_k$ is Gaussian white noise with covariance $\mbf Q_k = \text{diag}([0.0005, 0.0005, 0.0025, 0.0025])$. A sensor located at the origin helps track the projectile by providing bearing angle and range observations at 5 Hz, according to Eq. \eqref{range} and \eqref{angle}. The measurement noise covariance matrix associated is $\mbf R = \text{diag}([10^{-5}, 10^{-6} ])$. The projectile starts its motion at the origin with a given velocity: its initial condition is $\hat {\mbf x}_0 = [0,0,1,12]^T$, which is the mean of the initial state Gaussian distribution with covariance matrix $\mbf P_0 = 0.01 \mbf I_4$, being $\mbf I_l$ identity matrix of dimensions $l$.

\begin{figure}[ht]
    \centering
    \includegraphics[width=0.48\textwidth]{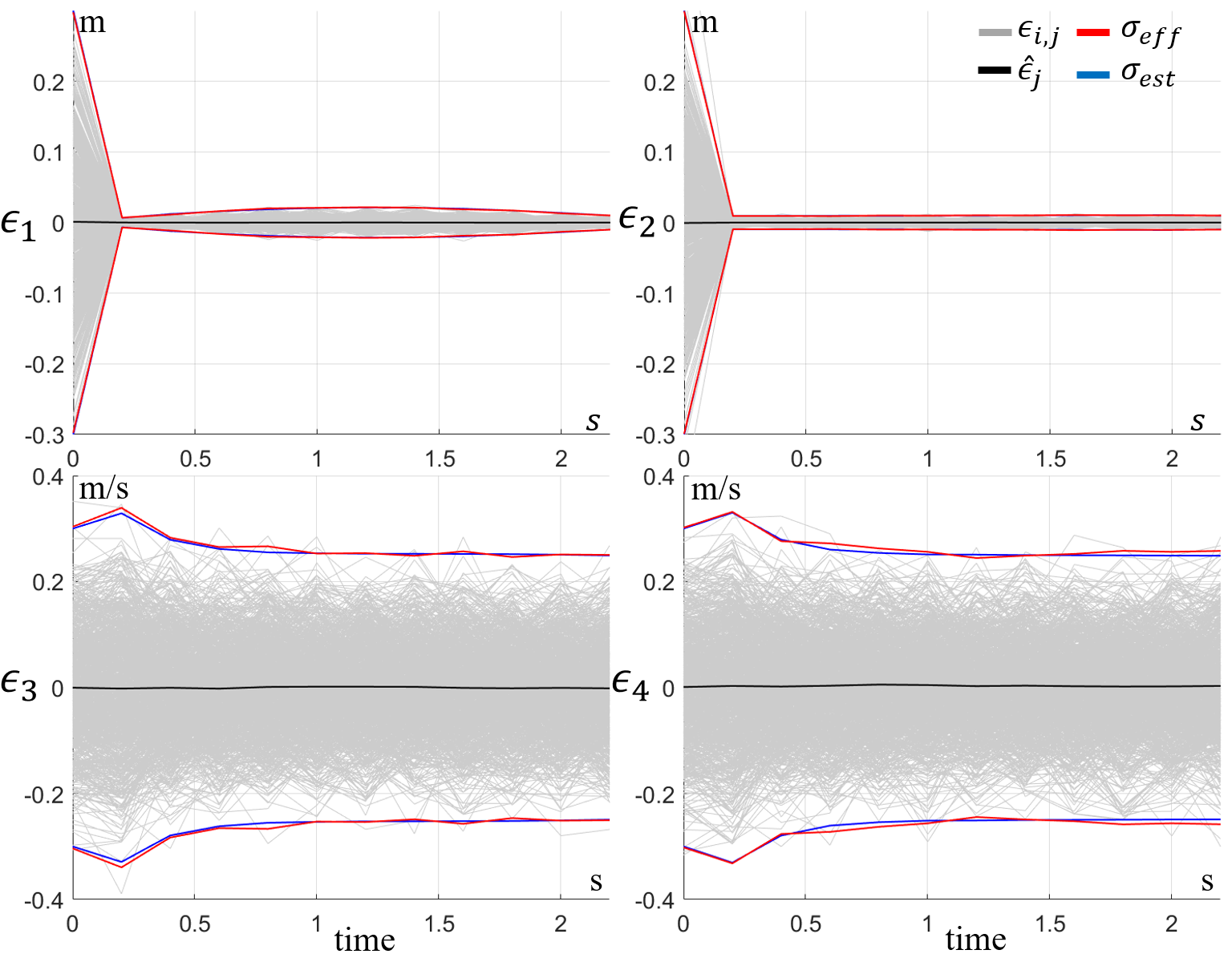}
    \caption{SPF-2 Projectile tracking error. }
    \label{track}
\end{figure}
Both SPF-1 and SPF-2 achieve an accurate tracking of the projectile. Figure \ref{track} shows a $N_{MC}=10^3$ runs Monte Carlo analysis performed with SPF-2. Similar results have been achieved with SPF-1. The error between the true and the estimated state of each of the four state components
\begin{equation}
    \epsilon_j = \hat x_j - x_{T,j} \quad \text{for} \ j = 1,\dots,4
\end{equation}
is reported in gray for each Monte Carlo run. The standard deviation (STD) boundaries are reported for both the effective, $3\boldsymbol\sigma_{eff}$, and the estimated covariance, $3\boldsymbol\sigma_{est}$. The effective STD, in red, is the true indicator of the spread of the error evaluated directly from the Monte Carlo runs, at each time step:
\begin{equation}
   \boldsymbol \sigma_{eff,j} = \sqrt{ \sum_{i=1}^{N_{MC}} (\epsilon_{j,i} - \hat{\epsilon}_j )^2 } \quad \text{for} \ j = 1,\dots,4
\end{equation}
where $\hat{\epsilon}_j$ is the error mean of the $j$th state. The estimated STD, instead, comes from the estimated covariance matrix of the filter at the competition of the measurement update, evaluated by square rooting the trace of the matrix:
\begin{equation}
   \boldsymbol \sigma_{est,j} = \sqrt{ \mbf P^{+}_{jj} } \quad \text{for} \ j = 1,\dots,4
\end{equation}
The effective STD and the estimated STD overlap, meaning that the SPF-2 can predict its own uncertainties correctly and is a consistent filter. Moreover, the mean of the errors, in black, is null for the whole duration of the simulation, confirming that the SPF-2 is an unbiased filter. It is essential to mention that these results have been obtained using only $10^3$ particles and 100 scout particles. Looking at the figure, the filter is able to improve the position estimate of the projectile thanks to the RADAR observations. However, the velocity is affected by high noises and SPF-2, not receiving any information regarding the velocity vector, can only estimate the velocity error to its steady state. In order to improve its estimate, we would have to provide the filter with observations of the velocity vector, similarly to the position, maybe adding a Doppler speed camera. The results of the filter's performance with only RADAR measurements have been reported since the non-square application is more challenging than the square one with the speed camera. 

\begin{figure}[ht]
    \centering
    \includegraphics[width=0.48\textwidth]{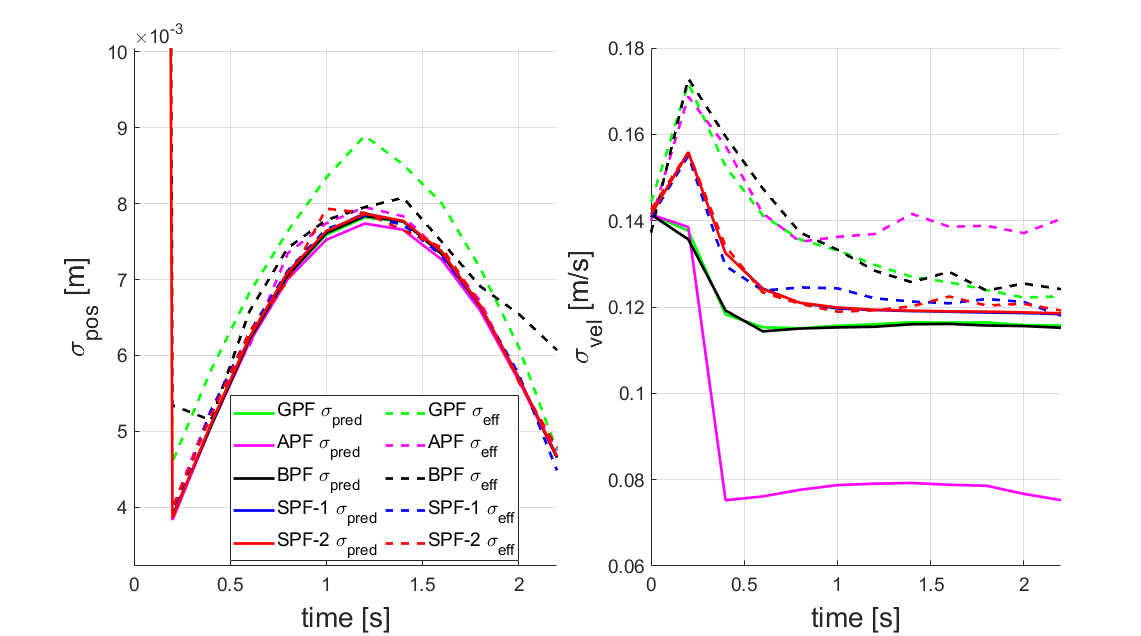}
    \caption{Standard deviation comparison among particle filters for the tracking problem. }
    \label{sig}
\end{figure}

The performance of SPF-2 has been compared to the other particle filters in terms of position and velocity STDs. In order to conduct an insightful comparison, the number of particles has been increased to $10^4$, as SPF was the only filter able to achieve convergence with such few particles as $10^3$. However, even with such modification, out of the $10^3$ Monte Carlo simulations, the Gaussian PF (GPF) still had 12 failed runs. Figure \ref{sig} shows the difference between the effective error, the dashed lines corresponding to how accurate each filter is, and the estimated error, the continuous line that represents how each filter estimates its accuracy. While the position is similar for all the filters, the velocity STD shows some differences. The position standard deviation has been zoomed in from the initial condition to show the parabolic behavior typical of the projectile motion, where it is harder to estimate its position at the instant of maximum altitude due to the high winds. First, the BPF and the GPF behave similarly, with a small overconfidence in their capability, as the effective covariance is larger than the estimated one. The Auxiliary PF (APF) has the biggest overconfidence, as its estimated STD is much smaller than the effective one. The APF uses an auxiliary variable to bring back information from the measurement onto the state PDF before propagation. Its performance is, therefore, heavily affected by the large process noise. Lastly, both the SPF-1 and SPF-2  have similar behavior. They are the only consistent filter methodology that accurately estimates its uncertainties, highlighted in the figure by the overlapping of the effective and estimated STD lines. 

\begin{figure}[ht]
    \centering
    \includegraphics[width=0.48\textwidth]{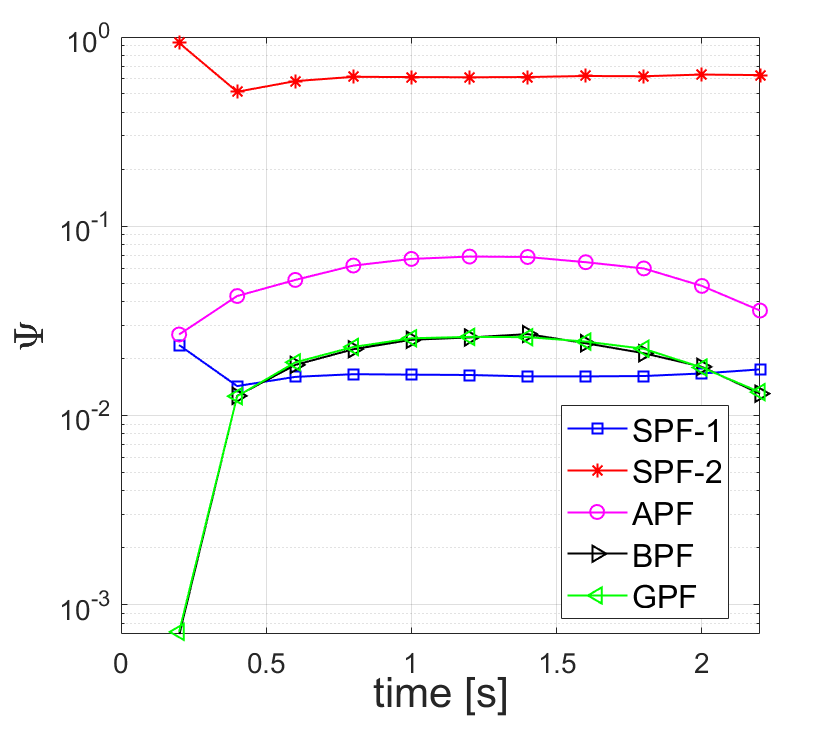}
    \caption{Efficiency comparison among particle filters for the tracking problem.  }
    \label{Psi}
\end{figure}

The Monte Carlo analysis also offers insights into the efficiency level of the filters by analyzing the parameter $\Psi$. Figure \ref{Psi} reports, at each time step, the percentage of effective particles of the filters with a logarithmic scale, evaluated as a mean. The SPF-2 efficiency level surpasses the other filters' level by more than a single order of magnitude. It is worth remembering that SPF-1 and SPF-2 are the only filters able to track the projectile with only 1000 particles. In SPF-1, the selection of drawing particles from a uniform distribution penalizes efficiency. This penalty becomes more prevalent as the number of dimensions increases, as the hypercube domain gains more corners where the distribution density is low. While it looks like the SPF-1 is the filter with the worst importance distribution, the results from Fig. \ref{sig} show that the SPF-1 centers its importance distribution onto the right region of the state space. A final comment is reserved for the shape of the curves in Fig. \ref{Psi}. The SPF-1 and SPF-2 lines are straight along the whole simulation: thanks to the map inversion, at each time step, the scout particles always highlight the same portion of prior, regardless of the specifics of the dynamics. On the contrary, the BPF, APF, and GPF share a parabolic trend. This change in efficiency is due to the projectile slowing down when reaching the peak of its trajectory and, therefore, being more subject to the influence of the process noise, as the velocity is at its lowest value. 

%%%%%%%%%%%%%%%%%%%%%%%%%%%%%%%%%%%%%%%%%%%%%%%%%%%%%%%%%%%%%%%%%%%%%%%%%%%%%%%%%%%%%%%%%
\subsection{Perturbed Orbit Determination}
The last application consists of determining the orbit of a spacecraft orbiting in Low Earth Orbit (LEO), affected by $J_2$ and $J_3$ gravitational perturbations. The equations of motion are derived using $\mbf{\ddot{r}} = \nabla U$ where:
\begin{align}
    U &= \dfrac{\mu}{r} + U_{J_2} + U_{J_3} \\
    U_{J_2} &= -\dfrac{3J_2\mu}{2r} \bigg(\dfrac{R_E}{r}\bigg)^2 \bigg(\sin^2(\phi)-\dfrac{1}{3}\bigg)^2\\
    U_{J_3} &= -\dfrac{J_3\mu}{2r} \bigg(\dfrac{R_E}{r}\bigg)^3 \bigg(5\sin^3(\phi)-3\sin(\phi)\bigg)^2
\end{align}
where $r$ is the satellite range, $\phi$ is the latitude, $J_2 = 0.0010826267$, $J_3 = -0.0000025327$, $\mu$ is Earth's gravitational parameters, and $R_E$ is Earth's radius. In this application, no process noise is considered. While the previous application highlighted the benefit of scout particles in systems affected by high process noises, this problem confirms that the SPF is robust even for systems that lack it. Process noise is critical for some particle filters to work correctly. For example, the BPF needs propagation noise to work properly to spread particles after resampling. Without noise, resampled particles that start from the same realization get propagated onto the same location. Therefore, the filter fails in the update step without the noise spread. 

The initial uncertainty for the orbit determination is of a Gaussian distribution with mean and covariance
\begin{equation}
    \hat {\mbf x}_0 = 
    \begin{bmatrix}
        -2012.151 \\ -381.450 \\ 6316.615 \\ 5.400366 \\ -5.916814 \\ 1.362965
    \end{bmatrix} 
    \quad
    \mbf P_0 = 
    \begin{bmatrix}
        30\ \mbf I_3 & \mbf 0_3\\
        \mbf 0_3 & 0.01\ \mbf I_3
    \end{bmatrix}
\end{equation}
Observations are acquired by a single sensor on the Earth's surface that provides range, elevation, and azimuth according to the equations: 
\begin{align}
    y_1 &= \sqrt{x_1^2+x_2^2+x_3^2} + \mu_1  \\ 
    y_2 &= \arctan \Big(\dfrac{x_2}{x_1}\Big)  + \mu_2  \\
    y_3 &= \arcsin \bigg(\dfrac{x_3}{\sqrt{x_1^2+x_2^2+x_3^2}}\bigg) + \mu_3
\end{align}
where measurement noises are assumed Gaussian with zero mean and 1 km STD for range and 1 arcsec for angle measurements. The difficulty of achieving a correct steady-state determination comes from the measurement acquisition frequency. Having a single sensor on the surface of the Earth leads to only two observation windows each day 12 hours apart, due to Earth's rotation. Whenever visible, it is assumed that the sensor can obtain measurements every 2 minutes for a total of 6 consecutive data acquisitions before the spacecraft goes out of sight for another half day. Therefore, this application is particularly challenging for particle filters as the state PDF noticeably increases between the long propagation and, simultaneously, there is no process noise. Indeed, the APF cannot be applied to this problem, and the BPF fails due to the lack of particle spread. 

\begin{figure*}[t]
    \centering
    \includegraphics[width=\textwidth]{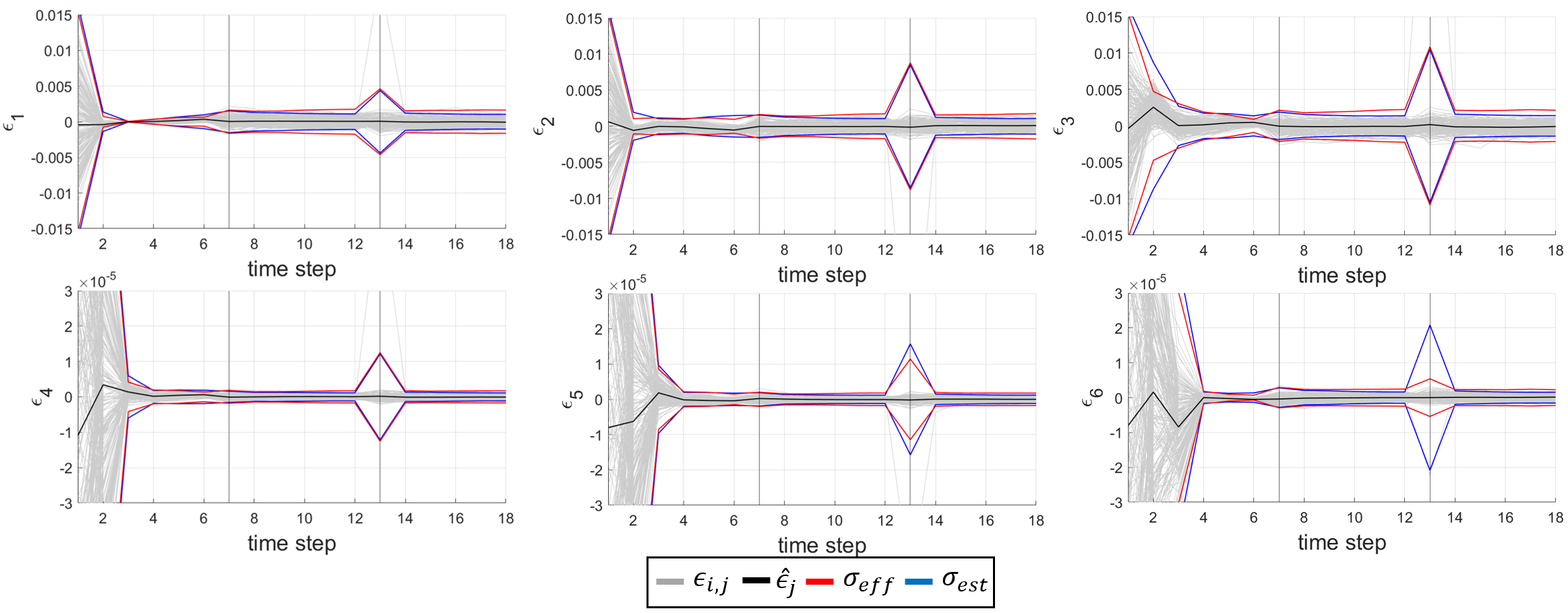}
    \caption{Orbit determination with SPF-2 and GPF. }
    \label{mcod}
\end{figure*}

A Monte Carlo analysis with 200 runs is reported in Fig. \ref{mcod}. The errors are shown with respect to the time steps rather than the time values since different time lengths are used in the simulation. Vertical lines are used to mark the 12-hour gap and the change of time window for the sensor. Likewise, for the previous application, effective and estimated covariances are reported for each state vector component. The filter shows a consistent behavior and the orbit of the satellite is tracked even after the long time propagations. Out of all the runs, only a single one spikes with a high error, occurring at a satellite's first sight. Nevertheless, using scouting particles, the filter overcomes the issue, bringing the error back to convergence levels. In the figure, the velocity errors have been zoomed in from the large initial uncertainties to show convergence. 

\begin{figure}[ht]
    \centering
    \includegraphics[width=0.48\textwidth]{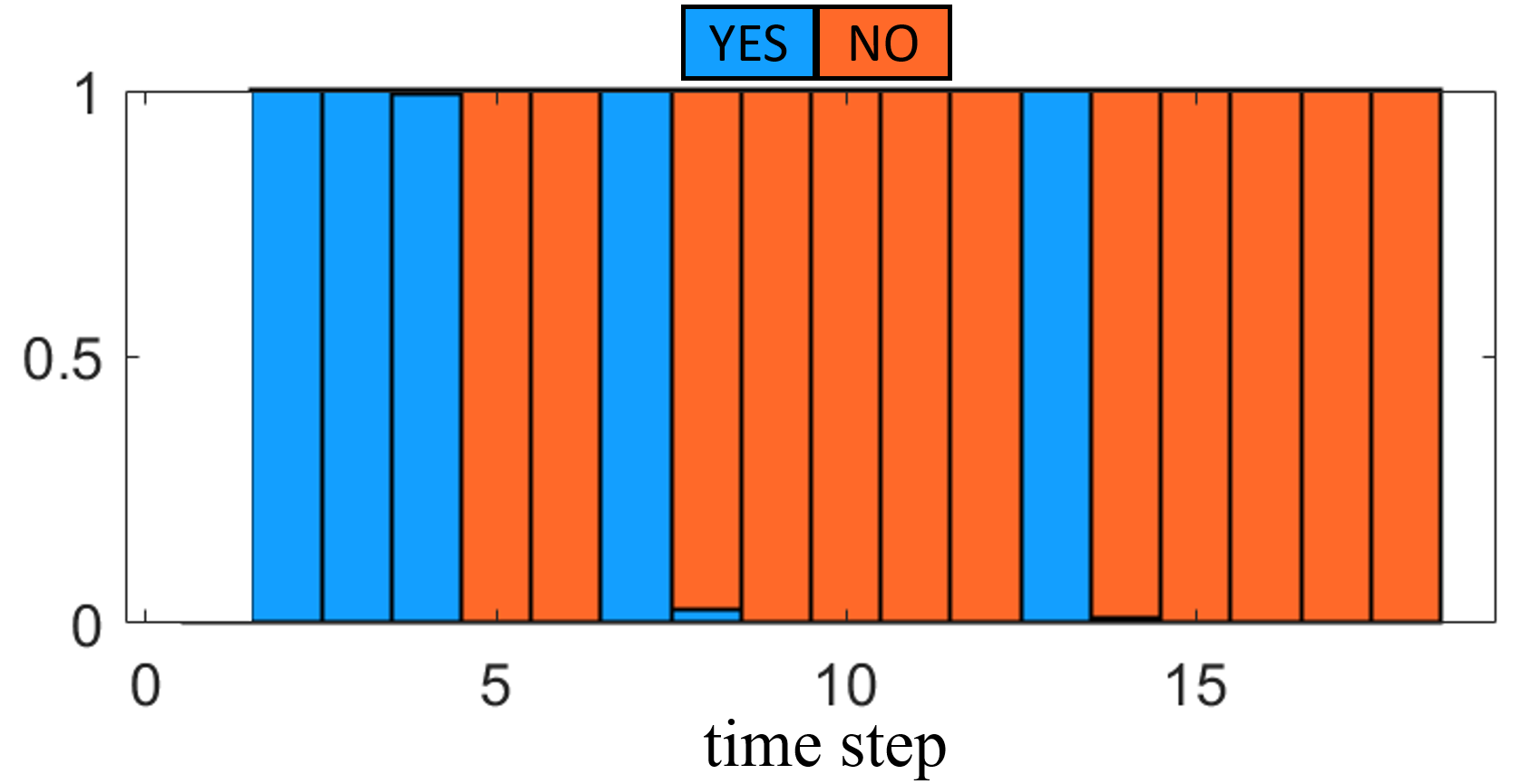}
    \caption{Scout particle measurement update selection.  }
    \label{which}
\end{figure}
The Monte Carlo analysis assesses the robustness of the new methodology where the other filters fail, and it shows that the scouting mechanism can be added to the existing techniques to strengthen common practices. The benefits of scouting are underlined in Fig. \ref{which}. After propagation, the filter can either apply SIS using the scouted distribution or the classic GPF update. The idea is that the scouting particles tell the filter which importance distribution is sufficient to sample from. When the prediction of the prior is poor, the scout particles inform the algorithm which region has high density, otherwise, when the scout covariance matrix has similar dimensions as the predicted one, the normal GPF update is performed. Thus, a comparison between $\mbf P^-_{k+1}$ and $\mbf P_q$ is computed at each time step, where the update technique connected to the covariance matrix with the smaller Frobenius norm is selected. Figure \ref{which} shows how many times, during the Monte Carlo analysis, the scout filter updated has been selected at each time step. The figure shows that the SPF-2 is fundamental to achieving convergence and having Initial Orbit Determination (IOD): the first three updates are almost entirely scout updates. When the posterior PDF becomes smaller, with a small error covariance after a few consecutive updates, the filter can rely on the GPF algorithm. However, whenever the sensor loses the satellite and has to wait 12 hours before acquiring a new observation, the filter adopts the scouting update to achieve an accurate estimate, as the GPF importance distribution is too widely spread. Thus, time steps 7 and 13 are scout updates for all simulations. In a few cases where the observation is particularly bad, the scout update is also needed in the following step (8 and 14) before the filter settles down and the complete GPF algorithm is adopted for the remaining measurement updates coming 2 minutes apart.

%%%%%%%%%%%%%%%%%%%%%%%%%%%%%%%%%%%%%%%%%%%%%%%%%%%%%%%%%%%%%%%%%%%%%%%%%%%%%%%%%%%%%%%%%
\subsection{Bimodal Distribution}
The last numerical application concerns a bimodal problem commonly used to test particle filters \cite{p4}. The discrete scalar system is governed by the following set of equations 
\begin{align}
    x_{k+1} &= \dfrac{1}{2}x_k + \dfrac{25x_k}{1+x_k^2} +8\cos(1.2k) + \nu_k \\
    y_k &= \dfrac{x_k^2}{20}+\mu_k
\end{align}
where $\nu_k$ is a zero mean Gaussian process noise with covariance $Q_k = 10$, and $\mu_k$ is zero mean Gaussian measurement noise with covariance $R_k = 1$. After inspecting the measurement equation, it appears evident that it gives no information regarding the sign of the state at the current time step, making the posterior PDF a bimodal distribution. Therefore, when inverting the measurement equation to provide the importance distribution with information regarding the likelihood, a bifurcation highlights two separate regions to sample from. Indeed, the scout particles are mapped back to both their possible parent priors, and the importance distribution reflects such region. 

\begin{figure}[ht]
    \centering
    \includegraphics[width=0.49\textwidth]{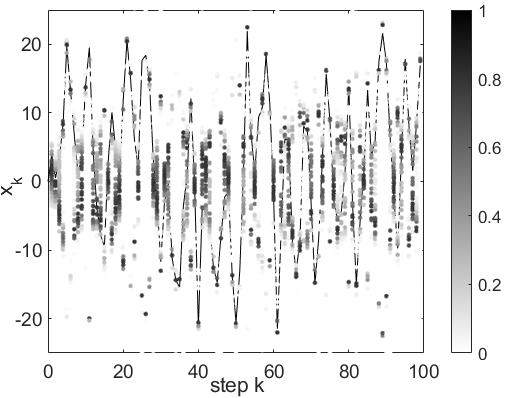}
    \caption{Evolution of probability density for the Scout Particle Filter with the true state (continuous line). }
    \label{shade}
\end{figure}

\begin{figure*}
    \centering
    \includegraphics[width=\textwidth]{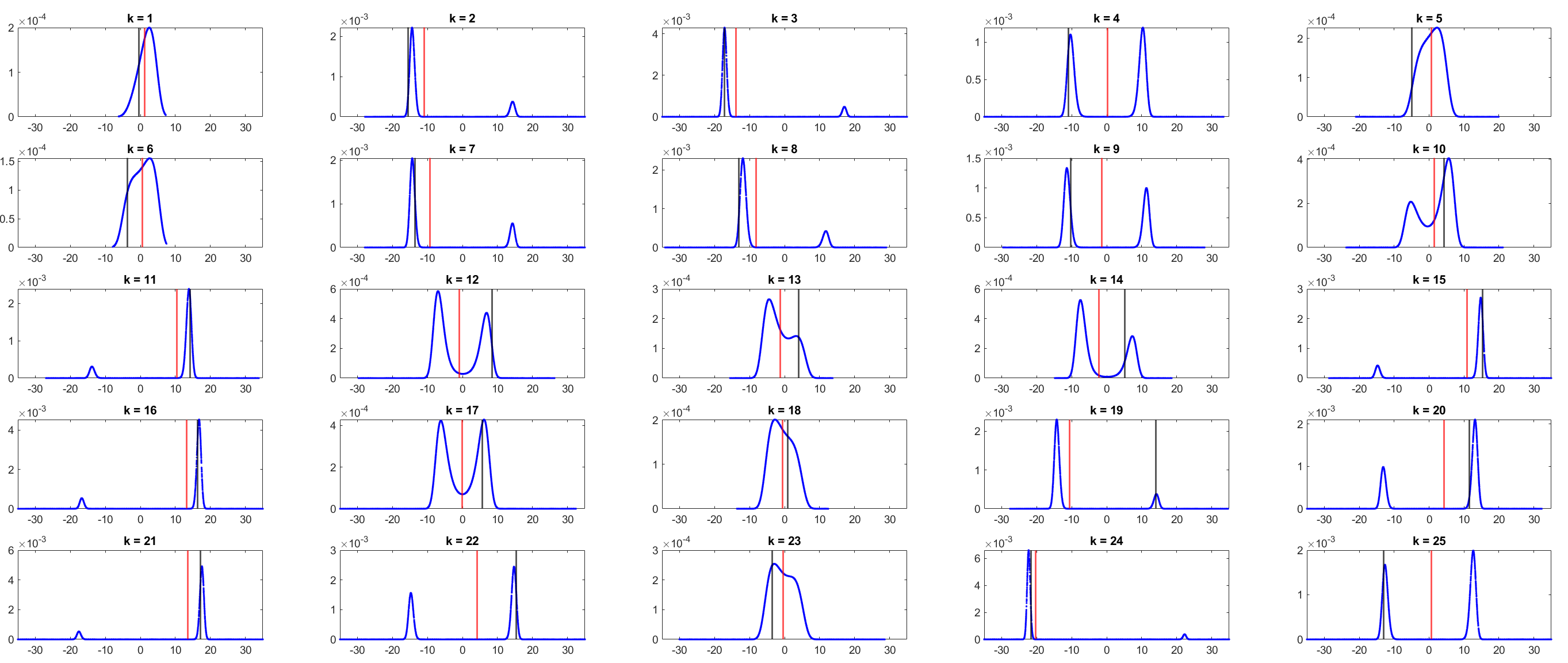}
    \caption{Posteriror approximation of the state for the first 25 steps of the simulation with true value (black) and estimate (red).}
    \label{postpdf}
\end{figure*}

The SPF has been applied to the system with 50 particles \cite{p4} and 20 scout particles. The representation of Fig. \ref{shade} reports the evolution of the probability density by assigning a grey scale to the particle weights, as well as the truth history of the state as a continuous black line. The selected representation follows the analysis reported in \cite{p4}, such that the reader can directly compare the SPF performance with the SIR-PF, the APF, the Regularized-PF, the GridBased-PF, and the Likelihood-PF, reported in the reference. Figure \ref{shade}  shows that the SPF can track the state of the system without neglecting the possibility of it belonging to the other mode of the posterior. 

In fact, state tracking loses importance for such a bimodal application as it becomes critical to estimate the posterior PDF correctly \cite{p4}. Consequently, rather than reporting the RMSE of the filter, the posterior distributions for the first 25 time steps of the simulation are reported in Fig. \ref{postpdf}. The figure shows, in blue, the location and relative weight of the particles, alongside their mean in red, and the true state, in black, as vertical lines. After analyzing highly bimodal steps such as the 9th, 17th, and 25th, it becomes evident how a correct estimation of the distribution is much more reliable than the mere state estimate. The SPF has predicted the state PDF correctly, accounting for the double mode of the function. 

This application compares the SPF with additional particle filters commonly used in the state of the art \cite{p4} and shows its additional capability of tracking bimodal systems.

%%%%%%%%%%%%%%%%%%%%%%%%%%%%%%%%%%%%%%%%%%%%%%%%%%%%%%%%%%%%%%%%%%%%%%%%%%%%%%%%%%%%%%%%%
%%%%%%%%%%%%%%%%%%%%%%%%%%%%%%%%%%%%%%%%%%%%%%%%%%%%%%%%%%%%%%%%%%%%%%%%%%%%%%%%%%%%%%%%%
%%%%%%%%%%%%%%%%%%%%%%%%%%%%%%%%%%%%%%%%%%%%%%%%%%%%%%%%%%%%%%%%%%%%%%%%%%%%%%%%%%%%%%%%%
\section{Conclusion} \label{concl}
The paper presented a novel way to select the importance distribution for SIS using Taylor polynomials. The new techniques are particularly effective when there is a poor knowledge of the state prior PDF, but it has the limitation of requiring functions to be differentiable. Nevertheless, numerical applications have shown the robustness and utility of the scouting mechanism in different applications. Indeed, the Scout Particle Filter achieved better performance than other common nonlinear estimators both on problems with high noises and long propagation times. 

The introduction of scout particles does not restrict the new methodology to any techniques. The scouting mechanism can be added and merged with other common particle filters without revolutionizing their overall structure. The results presented in the paper have shown that by doing so, the efficiency of the filters drastically increases, such that it is sufficient to draw fewer particles to achieve the same level of accuracy, reducing the computational burden. 

The SPF is able to achieve such results thanks to the polynomial map inversion of its polynomials. The creation of the map is the process that requires the most attention, as it directly influences the performance of the filter. Future work is dedicated to improving the map accuracy and expanding its capabilities to connect state distribution among multiple times and domains.

%%%%%%%%%%%%%%%%%%%%%%%%%%%%%%%%%%%%%%%%%%%%%%%%%%%%%%%%%%%%%%%%%%%%%%%%%%%%%%%%%%%%%%%%%
%%%%%%%%%%%%%%%%%%%%%%%%%%%%%%%%%%%%%%%%%%%%%%%%%%%%%%%%%%%%%%%%%%%%%%%%%%%%%%%%%%%%%%%%%
%%%%%%%%%%%%%%%%%%%%%%%%%%%%%%%%%%%%%%%%%%%%%%%%%%%%%%%%%%%%%%%%%%%%%%%%%%%%%%%%%%%%%%%%%
\section{Appendix}
Probability density functions are usually defined with the exponential function, e.g., exponential, gamma, chi-squared, and Gaussian distributions. Consequently, numerical issues can emerge when evaluating the probability of an outcome (particle) residing in a low-density region due to the exponential evaluation giving inaccurate results. In particle filtering, samples far away from the mean might get assigned null weights instead of the correct value. This appendix offers a different weight normalization methodology that applies the exponential on the difference of the exponents of particles rather than on the exponent itself. 

Consider a distribution embedded with an exponential. Let us define with $w_i$ the normalized weight of the $i$th particle that it is desired to calculate, with $a_i$ the weight of the particle prior normalization, and with $b_i$ the argument of the exponent, such that $a_i = C \exp(b_i)$, with $C$ the normalizing factor of the distribution. Given $N$ total particles, the normalized weight is
\begin{equation}
    w_i = \dfrac{a_i}{\sum_{j = 1}^N a_j}
\end{equation}
Inverting the equation, we obtain
\begin{align}
    \dfrac{1}{w_i} &= \dfrac{\sum_{j = 1}^N a_j}{a_i} = \sum_{j = 1}^N \dfrac{a_j}{a_i} = \sum_{j = 1}^N \dfrac{\exp(b_j)}{\exp(b_i)} = \sum_{j = 1}^N \exp(b_j-b_i) 
\end{align}
and inverting again
\begin{align}
    w_i &= \Bigg(\sum_{j = 1}^N \exp(b_j-b_i) \Bigg)^{-1}
\end{align}
It can be noted how the new normalization works only with the difference among the particles' location, without requiring the calculation of the exponential of their absolute position. This formulation is particularly useful in the DA framework, where the location of the samples is expressed with the deviation vector around a well-defined center. 

\begin{comment}

\end{comment}

\bibliographystyle{ieeetr}        % Include this if you use BibTeX 
\bibliography{autosam}

\begin{IEEEbiography}[{\includegraphics[width=1in,height=1.25in,clip,keepaspectratio]{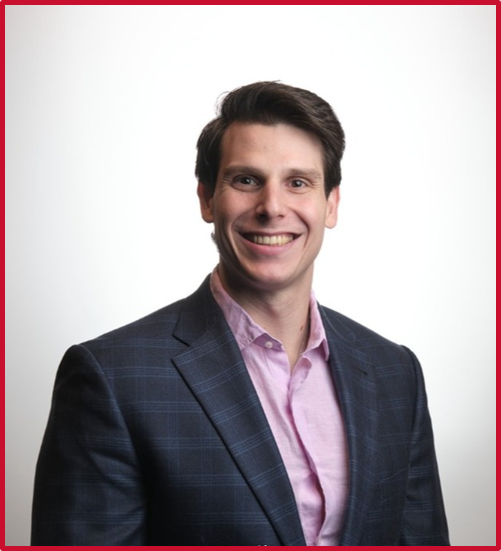}}]{\textbf{Simone Servadio}} (Member, IEEE)  received his Ph.D. degree in aerospace engineering from the University of Texas at Austin, Austin, TX, USA, in 2021. He is currently an Assistant Professor in the Department of Aerospace Engineering at the Iowa State University (ISU), Ames, IA. Before joining ISU, he worked as a postdoctoral associate at the Massachusetts Institute of Technology, Cambridge, MA, USA. His research interests include nonlinear estimation, autonomous navigation, optimal control, and space domain awareness. 
\end{IEEEbiography}

\end{document}